\documentclass[sigconf]{acmart}

\usepackage{acmart-taps}

\usepackage{graphicx}
\usepackage{fancyvrb}
\usepackage{xspace}
\usepackage[normalem]{ulem}
\usepackage[font={small,bf,sf}, textfont={small,sf}]{caption}
\usepackage{cleveref}
\usepackage{array}
\usepackage{longtable}
\usepackage{setspace}

\usepackage[formats]{listings}

\usepackage{pifont}

\usepackage{siunitx}

\lstdefineformat{C}{%
\{=\newline\string\newline\indent,%
\}=[;]\newline\noindent\string\newline,%
\};=\newline\noindent\string\newline,%
;=[\ ]\string\space}

\usepackage{booktabs}
\usepackage{multirow}
\usepackage{pifont}

\usepackage{balance}
\usepackage{scalefnt}
\usepackage{upquote}

\usepackage{soul}

\makeatletter
\patchcmd{\SOUL@ulunderline}{\dimen@}{\SOUL@dimen}{}{}
\patchcmd{\SOUL@ulunderline}{\dimen@}{\SOUL@dimen}{}{}
\patchcmd{\SOUL@ulunderline}{\dimen@}{\SOUL@dimen}{}{}
\newdimen\SOUL@dimen
\makeatother

\usepackage{marginnote}
\usepackage{comment}
\usepackage{footnote}

\usepackage{enumitem}

\usepackage{subcaption}

\usepackage{algorithm}
\usepackage{algorithmic}

\newlength{\gapspace}

\newcommand{\gap}[1][0cm]{\setlength{\gapspace}{#1}\vspace{.5\gapspace}\textbf{\color{red}[...]}\vspace{.5\gapspace}\xspace}

\definecolor{prpcol}{rgb}{0.9,0.2,0.2}

\newcommand{\noteprp}[1]{{\color{prpcol}\textbf{[\textsc{prp:} #1]}}}

\definecolor{ajpcol}{rgb}{0.2,0.7,0.2}

\soulregister\cite7
\soulregister\ref7
\soulregister\pageref7
\soulregister\eg7
\soulregister\unit7
\soulregister\noteprp7
\soulregister\gap7
\soulregister\mypar7
\soulregister\tinyskip7
\soulregister\code7
\soulregister\footnote7
\soulregister\notemm1
\soulregister\todo1

\sethlcolor{Yellow!20}

\newcommand{\ie}{i.e.,~}
\newcommand{\eg}{e.g.,~}
\newcommand{\etal}{et al.~}

\newcommand{\F}{Fig.\hspace{2pt}}
\newcommand{\T}{Tab.\hspace{2pt}}

\renewcommand{\unit}[1]{\mbox{\hspace{2pt}#1}\xspace}

\colorlet{punct}{red!60!black}
\definecolor{background}{HTML}{EEEEEE}
\definecolor{delim}{RGB}{20,105,176}
\colorlet{numb}{magenta!60!black}

\lstdefinelanguage{json}{
    basicstyle=\ttfamily\footnotesize,
    numbers=left,
    numberstyle=\scriptsize,
    stepnumber=1,
    numbersep=2pt,
    xleftmargin=2em,                    
    showstringspaces=false,
    breaklines=true,
    frame=lines,
    backgroundcolor=\color{background},
    literate=
     *{0}{{{\color{numb}0}}}{1}
      {1}{{{\color{numb}1}}}{1}
      {2}{{{\color{numb}2}}}{1}
      {3}{{{\color{numb}3}}}{1}
      {4}{{{\color{numb}4}}}{1}
      {5}{{{\color{numb}5}}}{1}
      {6}{{{\color{numb}6}}}{1}
      {7}{{{\color{numb}7}}}{1}
      {8}{{{\color{numb}8}}}{1}
      {9}{{{\color{numb}9}}}{1}
      {:}{{{\color{punct}{:}}}}{1}
      {,}{{{\color{punct}{,}}}}{1}
      {\{}{{{\color{delim}{\{}}}}{1}
      {\}}{{{\color{delim}{\}}}}}{1}
      {[}{{{\color{delim}{[}}}}{1}
      {]}{{{\color{delim}{]}}}}{1},
}

\newenvironment{myitemize}{%
\begin{itemize}[leftmargin=1em, itemsep=.1em, parsep=.1em, topsep=.1em,
    partopsep=.1em]}
{\end{itemize}}

\newenvironment{myenumerate}{%
\begin{enumerate}[leftmargin=1.50em, itemsep=0em, parsep=0em, topsep=.1em,
    partopsep=.1em]}
{\end{enumerate}}

\newenvironment{structure*}{\color{blue}\begin{myenumerate}}{\end{myenumerate}}

\newcommand{\tinyskip}{\vspace{3pt}}

\newcommand{\mypar}[1]{\tinyskip\noindent\textbf{#1.}\xspace}
\newcommand{\myparr}[1]{\tinyskip\noindent\textbf{#1}\xspace}

\usepackage{tikz}
\newcommand*\myc[1]{%
\scalebox{0.78}{\begin{tikzpicture}[baseline=-3pt]
  \node[draw,circle,inner sep=0.5pt, fill=black] {\textcolor{white}{\textsf{\textbf{#1}}}};
\end{tikzpicture}}}

\InputIfFileExists{locals-style}{}{}

\lstdefinestyle{tightjson}{
  language=json,
  numbers=left, firstnumber=1,
  frame=lines, framerule=0.3pt,
  framesep=2pt,                 
  xleftmargin=0pt, xrightmargin=0pt,
  framexleftmargin=0pt, framexrightmargin=0pt,
  numbersep=5pt,                
  aboveskip=2pt, belowskip=0pt, 
  showstringspaces=false
}

\usepackage{framed}

\newcommand{\sys}{\textsc{ExclaveFL}\xspace}
\newcommand{\NVFlareNoSec}{\textsf{NV\-Flare (No security)}\xspace}
\newcommand{\NVFlareCVM}{\textsf{NVFlare (CVM)}\xspace}

\copyrightyear{2026}
\acmYear{2026}
\setcopyright{cc}
\setcctype{by}
\acmConference[SoCC'26]{ACM Symposium on Cloud Computing}{November 18--20, 2026}{Singapore, Singapore}
\acmBooktitle{ACM Symposium on Cloud Computing (SoCC'26), November 18--20, 2026, Singapore, Singapore}
\acmDOI{10.1145/3815789.3827953}
\acmISBN{979-8-4007-2735-1/2026/11}


\begin{document}


\title[\sys{}: Verifiable Claims for Federated Learning using Exclaves]{\sys{}: Verifiable Claims for Federated Learning using Exclaves}

\author{Jinnan Guo}
\affiliation{%
 \institution{Imperial College London}
 \city{London}
 \country{United Kingdom}
}
\email{jinnan.guo@imperial.ac.uk}

\author{Kapil Vaswani}
\affiliation{%
 \institution{Indian Institute of Science Bangalore}
 \institution{SPARC}
 \city{Bengaluru}
 \country{India}
}
\email{kapil.vaswani@sparc-labs.in}

\author{Andrew Paverd}
\affiliation{%
 \institution{Microsoft Security Response Center}
 \city{Cambridge}
 \country{United Kingdom}
}
\email{andrew.paverd@microsoft.com}

\author{Peter Pietzuch}
\affiliation{%
 \institution{Imperial College London}
 \city{London}
 \country{United Kingdom}
}
\email{prp@imperial.ac.uk}

\begin{abstract}

In federated learning~(FL), data providers jointly train a machine learning model without sharing training data. This makes it challenging to provide verifiable claims about the trained model, \eg related to the employed training data, data sanitization, or training algorithm---a malicious data provider can simply deviate from the correct training protocol without detection. Prior FL training systems have used trusted execution environments~(TEEs) to protect the training computation, relying on the confidentiality and integrity of TEEs. TEE confidentiality, however, has been shown to be vulnerable to various attacks, \eg via side channels.

We describe \sys, a system for training FL models that establishes verifiable claims about models through fine-grained runtime attestation proofs. Since such proofs only require integrity, \sys generates them using  \emph{exclaves}, which are integrity-only execution environments without software-managed secrets, making them immune to data leakage attacks. \sys uses exclaves to attest data transformations during FL training, and the runtime attestation proofs then form an attested dataflow graph of the entire training computation. The graph is checked by an auditor to ensure that the trained model satisfies claims, \eg about the use of data sanitization by data providers or aggregation by the model provider. Since exclaves require minor yet proprietary firmware changes, we evaluate a variant with similar performance but slightly weaker security guarantees on existing AMD SEV-SNP hardware. We show that \sys introduces less than 12\% overhead compared to unprotected FL model training.


\end{abstract}

\maketitle



\section{Introduction}
\label{sec:intro}

As machine learning~(ML) models are used in new and diverse settings, there is an increasing impetus towards providing information about how models were trained. For example, the EU AI Act~\cite{EUAIAct} requires providers of general-purpose ML models to provide \emph{``information on the data used for training, [...] including the type and provenance of data and curation methodologies (e.g. cleaning, filtering etc.), the number of data points, their scope and main characteristics; how the data was obtained and selected as well as all other measures to detect the unsuitability of data sources and methods to detect identifiable biases''}.
In response to these requirements, \emph{datasheets for datasets}~\cite{datasheetsdatasets} describe the composition of training datasets, and \emph{model cards}~\cite{modelcard} offer information about model training. Providers of open-source models, \eg Apertus~\cite{apertus}, publish the training datasets and code; providers of closed-source models also make claims: \eg the model card for OpenAI's GPT-5 describes how the training data was pre-processed to reduce personal information~\cite{gpt5}. Such claims are also relevant to model providers: \eg to demonstrate to regulators that a model has \emph{not} been trained on personal or copyrighted data, or to show that a model \emph{has} been trained on certain content, \eg under-represented languages~\cite{apertus}.

Currently, such claims about datasets and model training are made without proof. For closed-source models, it is impossible to verify claims when interacting with the model via an inference API. Even for open-source models, verifying that a given model is indeed the result of executing the provided training code on a given dataset would require repeating the entire training process, and the trained model may still differ due to non-determinism. Therefore, model claims are typically accepted based on reputational trust.

With \emph{federated learning}~(FL)~\cite{mcmahan2017communication, kairouz2021advances, jere2020taxonomy}, the situation becomes worse, because training is performed by (potentially many) \emph{data providers}, rather than a single model provider. Claims about models trained using FL thus require reputational trust in all data providers, as well as the model provider. Yet FL is still appealing in many settings because it allows data providers to retain control over their datasets. Its use has been proposed for applications ranging from training predictive language models on user input~\cite{hard2018federated, ramaswamy2019federated}, to training fraud detection models while protecting transaction data of competing financial institutions~\cite{yang2019ffd}.

In this paper, we explore whether it is possible \emph{to provide verifiable claims about how an FL model was trained}. We observe that this would require systems support during training to discover deviations from the agreed training protocol that would invalidate specific claims. For example, the system would have to discover if a data provider skips a pre-processing step that removes personal information from a dataset. There could be various reasons for such deviations, ranging from accidental (\eg bugs in the training code) to actively malicious (\eg skipping steps to reduce costs).

A strawman design for creating an FL training system with verifiable claims would be to execute the training computation inside \emph{trusted execution environments}~(TEEs), such as Intel SGX enclaves~\cite{arnautov2016scone} or AMD SEV-SNP confidential virtual machines~(VMs)~\cite{sev2020strengthening}. TEEs ensure that the integrity and confidentiality of computation are preserved by the hardware: TEE memory is isolated, and TEEs offer attestation proofs, which contain a signed measurement of the TEE memory taken at deployment time. Indeed, prior designs of FL training systems have used TEEs to offer claims about trained models through attestation proofs~\cite{openfl, quoc21secfl, guo2024trustworthy}.

We argue, however, that TEEs are the wrong abstraction to create verifiable claims about FL model training for two reasons:

\tinyskip

\noindent
\textbf{(1)~Focus on static attestation proofs.} In TEE-based FL training systems~\cite{openfl, quoc21secfl, guo2024trustworthy}, TEEs take  memory measurements at deployment time to produce \emph{static} attestation proofs, which establish that the TEE runs the expected FL system and training code. A model provider can check each data provider's static attestation, \eg when establishing an initial authenticated network connection~\cite{openfl}. Such static attestation proofs, however, are hard to relate to claims about the final trained FL model: verifiable claims depend on how the training process was carried out at runtime, but a static attestation proof only captures the state at the start of the training process. 

\tinyskip

\noindent
\textbf{(2)~Reliance on confidentiality.} TEE implementations typically require confidentiality to provide unforgeable attestation proofs. Therefore, a distributed TEE-based FL training system that relies on attested TLS connections~\cite{guo2024trustworthy} would be vulnerable to an adversarial data provider that leaks the attestation-endorsed TLS session key from their TEE: this would allow them to deviate from the correct training protocol without detection by providing a correct attestation. TEEs strive for confidentiality by isolating TEE memory from external access. In practice, this is challenging due to advanced confidentiality attacks, \eg using side channels~\cite{foreshadow, zombieload, li2022systematic}.

\tinyskip

We conclude that FL training should instead rely on \emph{runtime} attestation proofs about \emph{data transformations} during FL training. The fact that data transformations, such as data sanitization~\cite{granite}, local training~\cite{kairouz2021advances}, poisoning checks~\cite{yin2018byzantine, flshield, gabrielli2023protecting}, and model aggregation~\cite{yin2018byzantine}, have been carried out faithfully can be related to claims about the final trained FL model. Importantly, these runtime attestation proofs only require guarantees about the \emph{integrity} of data transformations and not their confidentiality. Our approach can thus avoid the more challenging security model of TEEs, which must protect against both confidentiality and integrity attacks.

To realize the above approach, our work makes the following two contributions:

\myparr{(1)}~We describe a new abstraction called an \textbf{exclave}, which is a hardware-based \emph{integrity-only execution environment}. The security guarantees of exclaves do not depend on keeping data secret from the exclave operator. Unlike in TEE implementations, the integrity of exclaves is still preserved \emph{even if the exclave operator reads all data within the exclave}, either directly or through a side channel. Exclaves support attestation through a hardware-based mechanism (\eg a security co-processor~\cite{sev2020strengthening}) to measure code integrity and sign attestation proofs at runtime, which can include runtime data. The attestation signing keys are used and maintained by the hardware only, and are never accessible to software, making exclaves more resilient to data leakage attacks. We show how exclaves can be realized through minor modifications of existing TEE implementations.

\myparr{(2)}~With the help of exclaves, we describe the design and implementation of \textbf{\sys}, a system for executing FL training jobs that provides verifiable claims about trained FL models. \sys uses exclaves to generate runtime attestation proofs of all executed data transformations in the FL job: this covers the transformations performed by data providers and the aggregations performed by the model provider in each training round.
Each attestation proof, which we call an \emph{exclave data record}~(EDR), binds together (through hashes): (i)~the data inputs to a transformation; (ii)~the code performing a transformation; and (iii)~the resulting data output of a transformation. EDRs are collected and used to construct an end-to-end \emph{exclave dataflow graph}~(EDG) that then establishes global properties. The EDG provides a complete and integrity-protected representation of all data transformations that led to a final trained FL model. It can be used by an auditor to verify claims about the FL model, \eg that data pre-processing and model training have been carried out faithfully by data providers and that the model provider has aggregated the contributions of all data providers without bias.

\tinyskip

We create a prototype implementation\footnote{\url{https://github.com/microsoft/exclave-fl}} of \sys that extends an existing FL framework, NVFlare~\cite{roth2022nvidia}, to execute all FL data transformations inside exclaves. \sys achieves this without requiring changes to the framework implementation. Since our exclave design requires minor modifications to proprietary TEE firmware, we implement an exclave variant on AMD SEV-SNP using a virtual TPM~\cite{vtpm}, which offers similar performance but with slightly weaker security guarantees.

Our experiments show that \sys can produce verifiable claims with low overhead: the execution of an FL training job within exclaves and the generation of runtime attestation proofs reduce training throughput by less than 12\% compared to a baseline without any security guarantees.



\section{Verifiable Claims in Federated Learning}
\label{sec:background}

We provide background on federated learning~(\S\ref{sec:background:fl}) and discuss the challenges when providing verifiable claims about trained models~(\S\ref{sec:background:claims}). We then introduce trusted execution environments~(\S\ref{sec:background:tee}) and survey whether such approaches (and other techniques) can provide suitable verifiable claims~(\S\ref{sec:background:appoaches}). Finally, we describe the threat model and security requirements for our desired solution~(\S\ref{sec:background:threat_model}).

\subsection{Federated learning}
\label{sec:background:fl}

\begin{figure}[tb]
  \includegraphics[width=0.45\textwidth]{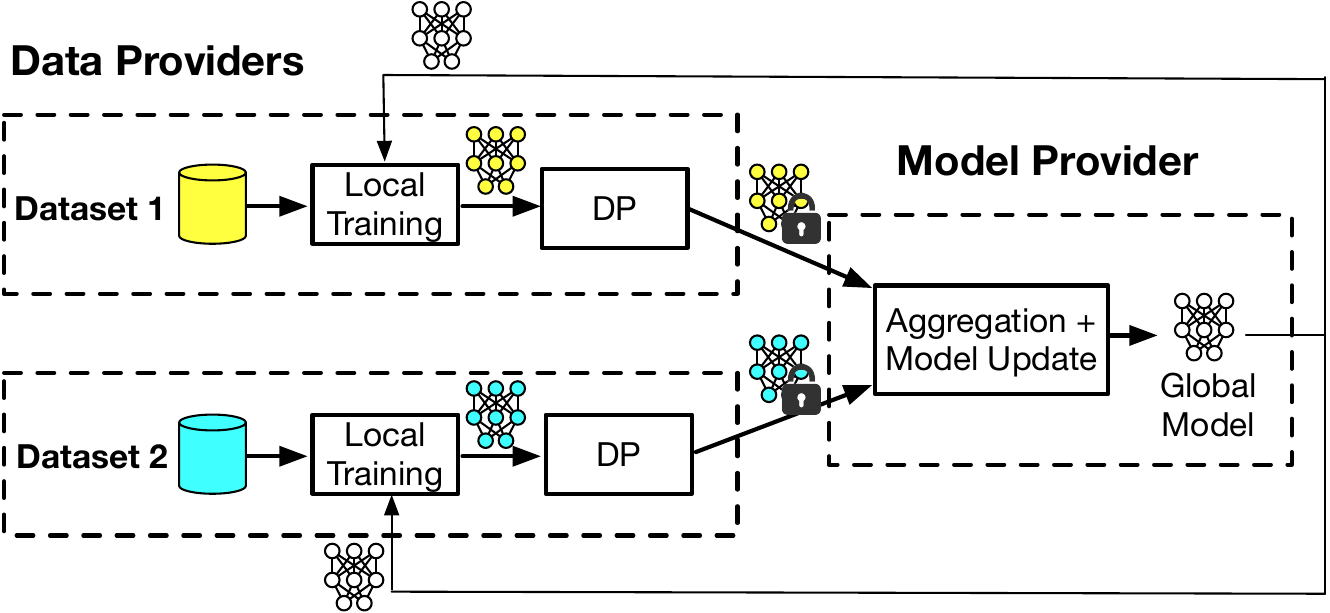}
  \caption{Overview of federated learning}
  \label{fig:fl_overview}
\end{figure}

Federated learning~(FL) trains machine learning~(ML) models while allowing data providers to retain control over their data. While FL applies to many scenarios, we consider the specific case of collaboration between multiple participants who wish to train models jointly using proprietary datasets, either for commercial gains, or to obtain models for their own use.
As shown in \F\ref{fig:fl_overview}, FL \emph{participants} jointly train a model: \emph{data providers} perform training locally using their own datasets and periodically send updates (either gradients or models) to a centralized \emph{model provider}. The model provider aggregates the updates and distributes updated model versions to the data providers. This process repeats for a fixed number of rounds until the model converges.
Techniques exist to protect the datasets or the quality of the trained FL model. For example, data providers may apply \emph{differential privacy}~(DP)~\cite{dwork2014algorithmic} to bound the contribution of individual data records in the training dataset. 
To ensure the quality of the trained FL model, FL jobs may include algorithmic approaches that detect or mitigate poisoning attacks~\cite{tolpegin2020data}, \eg using outlier detection~\cite{tolpegin2020data} or robust aggregation~\cite{blanchard2017machine, yin2018byzantine}.

\subsection{Claims about trained FL models}
\label{sec:background:claims}

It is an increasing requirement to provide information about aspects of the training process for ML models. Proposals for providing such information, now used in practice, include \emph{datasheets for datasets}~\cite{datasheetsdatasets} and \emph{model cards}~\cite{modelcard}. Analogous to their use in electronics, datasheets for datasets describe the composition and intended use of a training dataset; model cards perform a similar function for a trained model. Model cards should ideally include information about the training algorithm and the training datasets (\eg by referring to the datasets' datasheets), as well as details of any dataset pre-processing~(\eg to filter certain types of content).

In fully open-source models, the exact training code and dataset are made publicly available. In open-weight or closed-source models, the training code and dataset may be confidential, but model providers may still make public claims about the training code, dataset, and pre-processing. For example, the card for OpenAI's GPT-5 model claims that their data processing pipeline \emph{``includes rigorous filtering to maintain data quality and mitigate potential risks''} and uses \emph{``advanced data filtering processes to reduce personal information from training data''}~\cite{gpt5}. For all types of models, information about training may need to be provided to regulators. For example, Article~53 of the EU AI Act requires model providers to provide this information, \emph{``upon request, to the AI Office and the national competent authorities''}~\cite{EUAIAct}.

Making claims about FL training, \eg that particular datasets or attack mitigations have been used, is even more challenging. Claims about a trained FL model require trust in all participants who contributed to the training. There could be various reasons for any of these participants to deviate from the agreed training protocol, either unintentionally or deliberately. For example, a data provider may accidentally omit or misconfigure steps when pre-processing training data, leading to undesirable data being used for training. Alternatively, they may promise to train on a particular dataset but only train on a subset of this data to reduce costs.

It is an open challenge how to verify such claims about model training, even with access to the training code and datasets. For closed-source models that are only accessible via inference APIs, it is impossible to verify claims about model training; for open-source models, one could attempt to repeat the training process, but depending on the model size and training duration, this may require significant hardware resources, thus incurring substantial costs. Even if this were feasible, there is no guarantee that the trained model would match the original model due to the inherent non-determinism in the training process. Therefore, claims about model training typically have to be accepted based on \emph{reputational trust} in the entity making the claims. In an FL scenario, however, this reputational trust would have to extend to all participants, \ie the model provider and all data providers.

\subsection{Trusted execution environments}
\label{sec:background:tee}

Trusted execution environments~(TEEs)~\cite{sev2020strengthening, arnautov2016scone, cca24arm} try to ensure both the confidentiality and integrity of execution in a verifiable manner. Code and data in TEE memory are isolated, which prevents even privileged attackers, such as malicious administrators, operating system~(OS) kernels, or hypervisors, from accessing or tampering during execution. TEEs also support \emph{attestation}~\cite{buhren2019insecure} for external parties to verify a measurement taken over TEE memory, including both code and data. TEEs, including attestation mechanisms, are implemented in hardware, \eg relying on memory encryption, memory access checks, and hardware signatures for attestation.

TEEs exist at different granularities: ARM TrustZone~\cite{trustzone} provides a separate \emph{secure world}; Intel Software Guard Extensions~(SGX) offers isolation at a process granularity using \emph{enclaves}; more recently, \emph{VM-based} TEEs, such as AMD SEV-SNP~\cite{sev2020strengthening}, Intel TDX~\cite{mishra24tdx,cheng2023intel}, and ARM CCA~\cite{cca24arm,li2023enabling}, isolate entire VMs from the hypervisor and other privileged system software.

Hardware support for TEEs has also been integrated into GPUs. For example, Nvidia's H100 GPUs support a unified TEE that spans the CPU and GPU~\cite{dhanuskodi2023creating, gpublog}. In particular, it protects sensitive GPU state, including GPU memory and configuration registers, from external access. Communication between the CPU-based TEE and the GPU is encrypted and integrity-protected. Graphcore IPUs provide similar TEE functionality~\cite{vaswani2023confidential}.

\subsection{Approaches for verifiable claims}
\label{sec:background:appoaches}

Since TEEs protect computation at a hardware level and provide attestation proofs, an FL model provider could require all data providers to perform their model training inside TEEs and use remote attestation to demonstrate this in a provable manner. Existing approaches therefore propose hosting FL training jobs inside TEEs: CFL~\cite{guo2024trustworthy} demonstrates how to run FL frameworks in TEEs using SEV-SNP VMs coupled with NVIDIA Confidential GPUs; SecureFL~\cite{Kuznetsov2021SecureFL} is an FL training system that integrates SGX-based TEEs with ARM TrustZone on the edge to ensure privacy of both models and data; PPFL~\cite{mo_ppfl_2021} uses TEEs to prevent model and data providers from observing the model during training, which avoids membership inference attacks; SecFL~\cite{quoc21secfl} includes a policy manager to enforce attestation policies agreed between all clients; OpenFL~\cite{openfl} uses SGX-based TEEs during FL execution to prevent model leakage; GradSec~\cite{valerio21gradsec} targets FL privacy threats by protecting a subset of the model layers inside TEEs; CrowdGuard~\cite{crowdguard} mitigates model poisoning attacks by employing TEEs for client-side cross-validation on local models; Laminator~\cite{laminator} and Atlas~\cite{atlas} use TEEs to attest certain properties of centralized ML training and inference.

Using TEEs to provide verifiable claims about a trained FL model gives rise to two challenges: (1)~the attestation proofs offered by TEEs are based on measurements of the TEE memory at deployment time. Such \emph{static} attestation proofs include a hash over the FL job and framework code. Typically, they do not include a hash over the datasets, because the data may be unavailable at deployment time. This makes it challenging to relate such an attestation proof to any claims about the final FL model: for an auditor to verify claims linked to such proofs, they would have to re-run the attested FL training job and observe if they obtain the same final FL model.

In addition, (2)~TEE attestation relies on the confidentiality of TEE memory, because typical attestation implementations, \eg~Intel SGX quoting enclaves, rely on attestation keys stored in TEE memory. Since TEEs guarantee confidentiality, such an approach is sound, unless confidentiality is compromised. Here, TEEs have witnessed a long sequence of data leakage attacks that can disclose secrets from TEE memory, often by exploiting difficult-to-mitigate architectural and micro-architectural side channels~\cite{li2022systematic, wang2023pwrleak, spectre, li2021cipherleaks}. If such attacks are used to leak the attestation key from a TEE, the integrity of attestation proofs can be compromised by an adversary. This would enable the adversary to produce bogus attestations, which they could use as evidence for false model claims.

Such vulnerabilities extend beyond leaking the attestation key itself. Even for TEEs such as AMD~SEV-SNP, which keep the root attestation signing keys in a separate security co-processor, attestation mechanisms use the root keys to attest other keys, which then reside in TEE memory~(\eg TLS session keys). If such an attested key is extracted from TEE memory, an adversary could use it to send arbitrary model updates via the attested TLS connection while still appearing to be attested correctly.

Prior work has observed that guarantees about confidentiality and integrity could be separated in trusted execution. Sealed-glass proofs~(SGP)~\cite{sgp} use SGX-based TEEs as ``transparent enclaves'' to provide attestation proofs of computation over secret inputs, while allowing a host to inspect the TEE memory. Transparent enclaves, however, still assume that attestation keys remain secret: with SGX enclaves, the attestation key is held inside a separate quoting enclave~(QE) residing in TEE memory, which makes it vulnerable to side-channel attacks~\cite{foreshadow}.

Instead of relying on hardware mechanisms, other approaches have used algorithmic and cryptographic techniques to establish claims about trained models. \emph{Proof of learning} approaches~\cite{jia2021pol, abbaszadeh2024zero} build on techniques such as proof-of-work~(PoW)~\cite{bitcoin}, verified computation~\cite{safetynets}, and zero-knowledge proofs~\cite{zkp1, zkp2}. These approaches, however, either incur prohibitively high overheads or can be subverted~\cite{zhang2022adversarialPOL, fang2023broken}. For example, zero-knowledge proofs such as zkSNARK~\cite{zkp1} introduce orders-of-magnitude computational overheads~\cite{zkp2}; PoW-based mechanisms~\cite{pow} require the majority of the participants to follow the protocol correctly. This assumption does not hold in FL because participants are mutually distrusting.

\subsection{Threat model and security goals}
\label{sec:background:threat_model}

We assume that any participant in the FL training job (\ie a data provider or the centralized model provider), or a set of colluding participants, could be adversarial. This also includes the case in which honest participants have been compromised by an adversary.

\mypar{Adversary's goal} The adversary aims to deviate from the predefined FL training protocol in a way that modifies or influences the final FL model, \emph{and} for this deviation to remain undetectable by other participants and/or third-party auditors. 
This goal encompasses many specific attacks that participants could attempt: \eg an adversarial data provider may perform fewer training rounds to reduce resource costs, skip a pre-processing step that would otherwise filter out specific data, or alter their dataset during training to bias the model; similarly, an adversarial model provider may omit updates from specific data providers during aggregation to bias the model, or send different models to different data providers to infer information about a particular provider's private dataset~\cite{model_inconsistency}. More generally, any adversarial participant may want to deviate from the predefined FL training protocol by changing their training code, or modifying or falsifying data transferred during the FL training job.

\mypar{Adversary's capabilities} We assume a strong adversary with privileged access to a participant's system. This includes full control of the OS and hypervisor of a participant's machine, as well as the ability to intercept and modify all inbound and outbound network communication. In addition, the adversary can carry out side-channel attacks to leak memory contents, including TEE memory. We assume, however, that the adversary does not carry out advanced physical attacks against the hardware, such as physically extracting keys from a separate on-chip security co-processor, due to the high costs and expertise required.

\mypar{Security guarantees} \sys aims to ensure that the aforementioned adversary cannot deviate from the predefined FL training protocol, in a way that modifies or influences the final model, without detection. \sys thus enables any participant or third-party auditor to verify claims about the FL training process.
We define a \emph{verifiable claim} to be any statement about the FL training process that could have been made by a (hypothetical) trusted omniscient observer with full visibility of the training process. Claims could include statements about the executed code for training and aggregation, the number of participants and rounds, or any other runtime property of the FL job. It is also possible to make statements about the precise data used for training, but sometimes it is desirable for the data to remain private to the data providers. In such cases, statements can be made about a \emph{representation} of the data, \eg a blinded hash to confirm that the data has not been changed during FL training. In addition, claims may make statements about what pre-processing (\eg data filtering) has been performed.

\mypar{Security non-goals} Given our assumed adversary's capabilities, we cannot \emph{prevent} deviation from the FL training protocol, but only require detection. Similarly, we cannot guarantee the \emph{availability} of verifiable claims---there are various ways in which our strong adversary can prevent such claims from being made (\eg through a denial-of-service attack). 
In addition, we do not provide \emph{model confidentiality} from data providers. Instead, we assume FL scenarios in which the trained model is ultimately shared with all participants,~\eg a group of hospitals that is jointly training a model on patient data.
Finally, we do not aim to detect the poisoning of a training dataset \emph{before} training begins. There is a large body of related work on the detection or prevention of such attacks~\cite{threats_fl}. \sys supports such mitigations by (i)~ensuring that they have been applied correctly by the respective participants; and (ii)~allowing third parties to verify that they were in place during training.



\section{Exclaves}
\label{sec:exclaves}

We introduce \emph{exclaves}, a new abstraction that overcomes the challenges described above. We first explain the exclave concept~(\S\ref{sec:exclaves:concept}) and then describe how exclaves can be implemented~(\S\ref{sec:exclaves:impl}).

\subsection{Exclave concept}
\label{sec:exclaves:concept}

Current TEE implementations use attestation to establish a trust relationship between the TEE and a verifier. For example, an attested TLS network connection~\cite{fossati-tls-attestation-08} can prove that the verifier is communicating with a TEE and that all messages received via the TLS connection were sent by that TEE.
An adversary with the ability to leak data from a TEE (\eg through side channels), however, can undermine the security guarantees of current TEE implementations when used to secure FL training (see~\S\ref{sec:background:threat_model}). In the above example, the adversary could leak the session key of the attested TLS connection and masquerade as the TEE. In the presence of such an adversary, we therefore cannot rely on secrecy of data within the TEE.

\mypar{Exclaves} We introduce the \emph{exclave}, which is an \emph{integrity-only} execution environment that does not contain data that must be kept secret from an adversary. Similar to TEEs, an exclave ensures that code is executed without external interference, and that the data within the exclave can only be modified by code within the exclave. An exclave thus offers \emph{integrity} guarantees for its code and data, even against privileged software running on the same host.

Unlike TEEs, exclaves do not provide confidentiality guarantees---the only piece of data that must be kept secret is a single attestation signing key. As we explain in \S\ref{sec:exclaves:impl}, the attestation key can be held by a security co-processor, which is hardened against side-channel attacks by being physically separate~\cite{frazelle2019securing}. 
While TEEs can execute the same workloads as exclaves, the attestation mechanisms of TEEs assume confidentiality of TEE memory.

Code that executes inside an exclave can request the security co-processor to produce a signed attestation report, which contains (i)~a measurement of the exclave's code and data at deployment time and (ii)~user-defined data supplied by the application code inside the exclave at runtime. This allows the attestation report to prove that a particular claim was made by the exclave at runtime.

\mypar{Exclave Data Records~(EDRs)} The attestation report can be used to produce an unforgeable record of the \emph{data transformations} carried out by the exclave, \ie any computation that takes a set of data inputs and produces a set of data outputs. Exclaves can thus be used to attest individual data transformations at runtime. After performing a data transformation within an exclave, the exclave can produce a runtime attestation proof that binds the data transformation to the specific code and data of the transformation. We refer to such an attestation proof as an \emph{exclave data record}~(EDR).

More formally, an EDR is an existence proof that at least one execution of the specified code carried out the stated data transformation. In terms of its implementation, an EDR is a signed statement that contains (i)~a measurement of the exclave's code, defining the computation; and (ii)~a representation of the data transformation, containing two key/value structures: one for inputs and another for outputs. The keys uniquely identify inputs and outputs (\eg \texttt{\small{}model\_weights} or \texttt{\small{}training\_data}); the values are fixed-size cryptographic hashes of the input and output data. Analogous to TEEs, an exclave's code measurement is a hash of the code pages loaded at deployment time. Given access to the code, a verifier can thus recompute the hash and compare it against the measurement.

\subsection{Exclave implementation}
\label{sec:exclaves:impl}

An exclave can be implemented with minor modifications of a TEE implementation, such as AMD~SEV-SNP~\cite{sev2020strengthening}, by reusing its isolation and hardware-based attestation mechanisms. With AMD~SEV-SNP, exclaves can rely on the memory isolation enforced by the memory management unit~(MMU): during address translation, the MMU checks the access policy stored in the \emph{reverse map table}~(RMP), which is managed by the CPU and the \emph{platform security processor}~(PSP)~\cite{amd_kds, bagia2025close, sev2020strengthening}. The exclave uses the PSP to generate hardware-signed attestation reports at runtime. The signing key for attestation reports, \ie the \emph{versioned chip endorsement key}~(VCEK)~\cite{sev2020strengthening, buhren2019insecure}, is stored and managed by the PSP only.

Exclaves, however, must generate runtime attestation reports by passing data securely to the PSP, without relying on shared software secrets---any software secret would violate their secret-free nature. In AMD~SEV-SNP, each TEE maintains \emph{VM platform communication keys}~(VMPCKs) to protect communication with the PSP. An exclave implementation using AMD SEV-SNP must therefore protect PSP communication through an MMU access check instead of a shared secret: the exclave must use a special \emph{protected page}, accessible only by the exclave and the PSP, to create attestation reports.

An exclave implementation using AMD~SEV-SNP would thus require the following firmware changes: with exclaves as CVMs, a protected page could be implemented as a runtime-accessible analogue of SNP's existing \emph{secrets page}, which is RMP-protected, guest-owned, but PSP-writable only on provisioning. This change would require: (1)~defining a new guest-owned protected page, whose RMP entry has the same metadata as the secrets page (\eg exclave's guest ASID, validated-bit, etc.); (2)~extending the firmware with functionality for the PSP to read/write this protected page at runtime and not only at provisioning time; and (3)~recording this page type as \texttt{PAGE\_TYPE\_PROTECTED} and provisioning it during exclave creation (similar to secrets page provisioning).

With these changes, access to the protected page would be enforced by SNP's RMP check, enabling secret-free attestation. These firmware changes would be feasible, because they reuse the existing implementation of the secrets page protection. Moreover, they would not affect the verification of the CVM's initial state post-boot: the expected hash of the CVM firmware image would need to be updated, but this is the same as for a standard firmware update.

For other TEE implementations, achieving secret-free attestation would require the removal of privileged software components, such as \emph{security monitors}~(SMs), from the runtime attestation process. For example, the Keystone TEE~\cite{keystone} would require a secure hardware signer (\eg HSM/TPM~\cite{tpm}), making its SM an untrusted proxy for attestation requests to that hardware signer. Compared to other TEEs, AMD SEV-SNP already provides hardware-signed runtime proofs by the PSP, which matches an exclave's requirements.

Since we cannot modify the proprietary AMD SEV-SNP firmware, we emulate equivalent signing functionality for exclave attestation. As we describe in~\S\ref{sec:impl:attestation}, our exclave implementation on unmodified AMD~SEV-SNP hardware emulates the modified attestation approach using a \emph{virtual TPM} as a stand-in for the PSP's attestation functionality, which yields comparable performance.



\section{Exclave Federated Learning}
\label{sec:tfl}
\label{sec:tfl:design}

We describe the design of \emph{Exclave Federated Learning}~(\sys), a system that produces verifiable claims about an FL training process and the resulting model. \sys{} uses exclaves to produce runtime attestation proofs, in the form of EDRs, of all data transformations leading to a final trained FL model. The collected set of EDRs thus provides a trustworthy end-to-end record of the training process over all FL computation rounds. An auditor can use the EDRs to verify claims about the trained FL model~(see~\S\ref{sec:flex}).

The use of exclaves in an FL training system introduces two challenges: (i)~the integrity of the claims must not depend on the confidentiality of the FL training process, because, unlike TEEs, exclave execution is visible to FL participants; and (ii)~to keep the TCB small, the claims should only require the minimum amount of code to be integrity-protected (\ie executed inside exclaves). In particular, the claims should not rely on the integrity of the FL framework code, as long as the EDRs attest the data transformations to establish correctness.

\begin{figure}[tb]
  \includegraphics[width=0.48\textwidth]{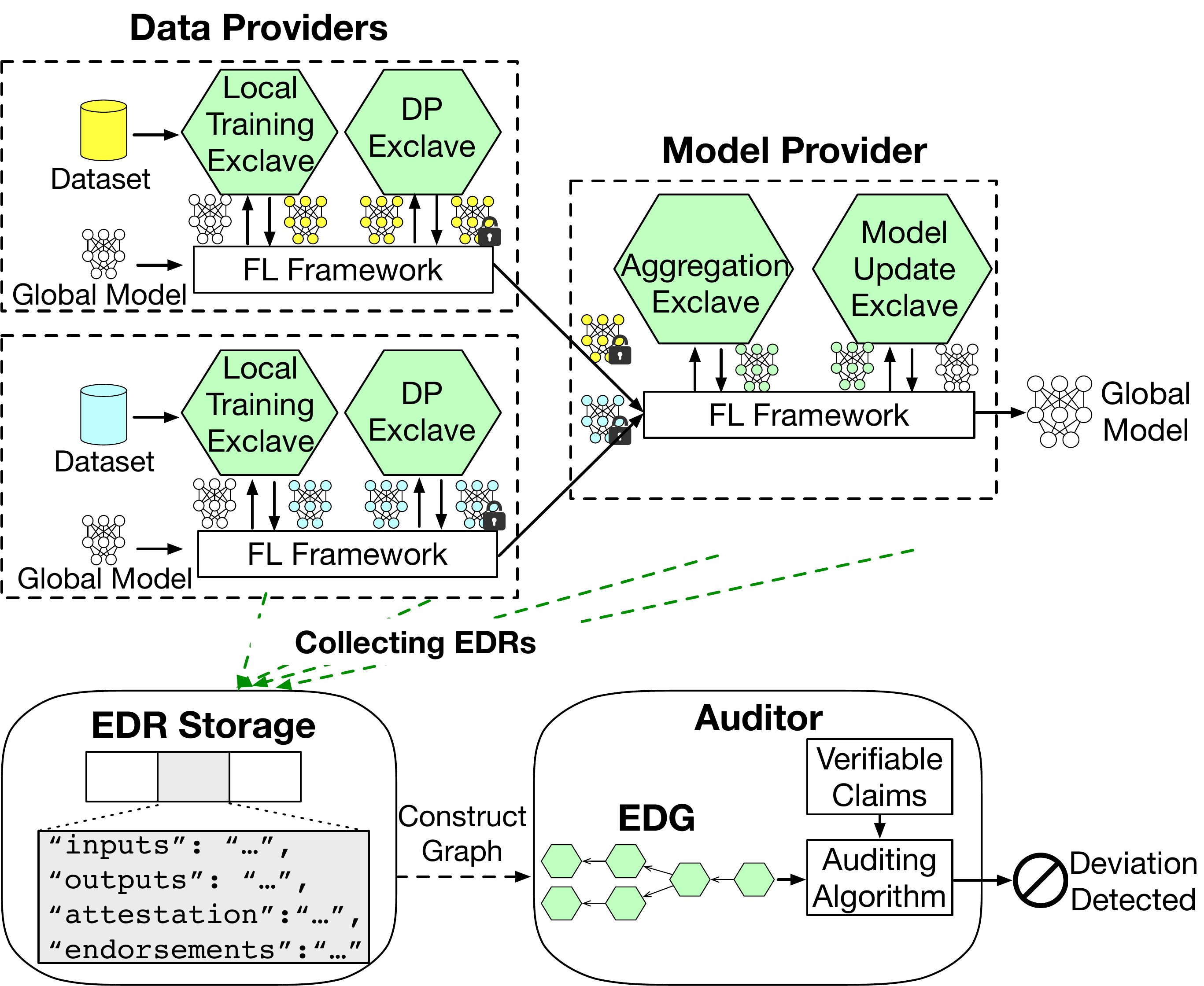}
  \caption{Design of \sys \textnormal{(Exclaves are shown in green.)}}
  \label{fig:workflow}
\end{figure}

As shown in \F\ref{fig:workflow}, \sys executes an FL training \emph{job} as a sequence of FL \emph{tasks} that define the data transformations. Data providers and the model provider execute tasks in each training round. An FL framework such as NVFlare~\cite{roth2022nvidia} provides the task implementations, orchestrates task execution and offers APIs for data movement between tasks. \sys then executes each task within an exclave, which ensures execution integrity and produces EDRs for each data transformation.

In an FL job, the tasks executed by data providers may include: (i)~a \emph{local training task} that takes a global FL model, reads a dataset from storage, computes a local model update, and returns the updated local model; and (ii)~a \emph{differential privacy~(DP) task} that inputs a local model update, adds DP noise to protect privacy, and outputs a DP-protected local model update. A model provider may execute: (iii)~an \emph{aggregation task} that receives DP-protected local model updates from data providers and outputs an aggregated model; and (iv)~a \emph{model update task} that converts the aggregated model to a global FL model. The updated global FL model is sent to the data providers ahead of the next training round. Note that the rest of the FL framework executes outside of exclaves, thus keeping the TCB limited to the data transformations that affect the FL model. 

Since \sys executes tasks inside exclaves, the produced EDRs create a trustworthy record of the executed data transformations. For example, an EDR produced by the aggregation task contains hashes of (i)~the local model updates received by the task in a round and (ii)~the computed aggregated model; an EDR produced by the model update task has hashes of (i)~the aggregated model, (ii)~the previous global model, and (iii)~the updated global model; and an EDR produced by the local training task contains hashes of (i)~the received global model, (ii)~the training dataset, and (iii)~the updated local model. If the FL job executes faithfully, the EDRs must show that the correct code for the task implementations was run. In addition, the hashes of the data inputs/outputs in the EDRs must match between EDRs produced by different tasks: \eg the output hash of the updated global model in the EDR generated by the model update task (executed by the model provider) must match the input hashes of the global model in the EDRs generated by the local training tasks (executed by the data providers).

When the local training task reads a dataset from storage, \sys must ensure that the input hash in the EDR remains valid. Therefore, a data provider must provide the training dataset as an integrity-protected disk image~\cite{dmverity}, protected by a Merkle tree~\cite{merkletree} to preserve the integrity of the blocks. When the dataset is accessed by the local training task in the exclave, \sys checks the dataset blocks against the Merkle tree. The root hash of the Merkle tree (combined with a random salt) serves as a non-revocable \emph{data commitment} to the training dataset, which is included in the EDR.

When exclaves produce EDRs, the EDRs are sent to an \emph{EDR storage service}. This service maintains all generated EDRs during job execution and makes them available for auditing. Note that the EDR storage service does not need to be trusted: since EDRs are integrity-protected by a cryptographic signature, any tampering is detectable. If individual EDRs are withheld by an adversary from the EDR storage service, this is revealed by the introduced inconsistencies between EDRs (see below). As we explain in \S\ref{sec:flex}, an \emph{auditor} can check the EDRs for mismatches and inconsistencies, which indicate a deviation from the faithful execution of the FL job and invalidate the claims about the trained FL model. The auditor can be an FL participant, a consumer of the trained FL model who needs to verify correct training, or a regulator who audits the training process.



\begin{figure}[t]
  \centering
  \includegraphics[width=0.3\textwidth, trim = {0 40 0 0}, clip]{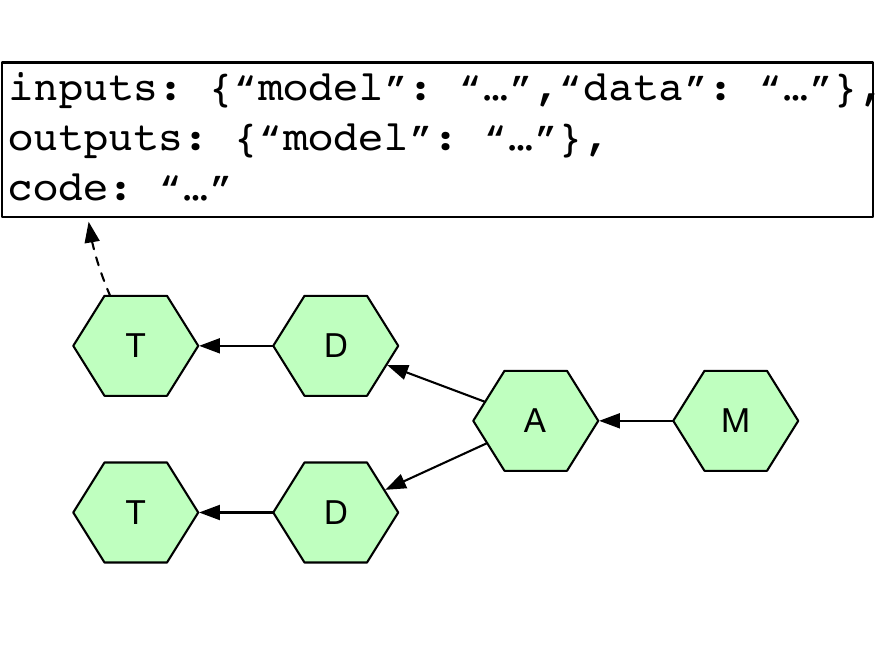}
  \caption{Example of an Exclave Dataflow Graph \normalfont{(Edges show reverse dataflows; nodes show \underline{T}raining, \underline{D}P, \underline{A}ggregation, and \underline{M}odel update tasks.)}}
  \label{fig:correct_edg}
\end{figure}

\section{Auditing of Verifiable Claims}
\label{sec:flex}
Auditing verifiable claims involves constructing a graph based on the EDRs~(\S\ref{sec:flex:edg}) and then verifying the claims against the graph~(\S\ref{sec:flex:auditing}).

\subsection{Constructing exclave dataflow graphs}
\label{sec:flex:edg}

The EDRs collected by the EDR storage service are used to construct a directed acyclic graph, which we refer to as the \emph{exclave dataflow graph}~(EDG). \F\ref{fig:correct_edg} shows a sample EDG: each vertex~$v \in V$ is an EDR, which is defined by the tuple $(\mathit{in}, \mathit{code}, \mathit{out})$, which refers to the data input, code, and data output hashes from that exclave execution, respectively. Each edge~$e \in E$ represents the (reversed) data flow between exclaves; it points from the downstream (consumer) vertex~$v_{y}$ to the upstream (producer) vertex~$v_{x}$:
\begin{equation}
E=\{(v_{y},v_{x}) \mid v_{y},v_{x} \in V \text{ and } v_{y}.\mathit{in} \cap v_{x}.\mathit{out} \neq \emptyset \}
\end{equation}

The edge label represents the data that flows between the two exclave executions. It is defined as the non-empty intersection of the outputs of $v_x$ and the inputs of $v_y$ (since exclaves may have multiple inputs and outputs for a single execution):
\begin{equation}
l(v_y,v_x) = v_y.in \cap v_x.out
\end{equation}

We define the $\mathit{EDG} = (V, E, l)$ as a collection of vertices~$V$, reversed dataflow edges~$E$, and the edge label function~$l$. For auditing, Alg.~\ref{algo:edg-generate} shows how \sys constructs an EDG from a collection of EDRs~$v$, which are attested as~$\sigma(v)$. For each EDR, \sys first verifies the attestation with the hardware manufacturer, \eg using AMD's Key Distribution Service~\cite{amd_kds}~(line~\ref{algo:edg-generate:verify}). If the verification succeeds, \sys adds the EDR to the EDG as a vertex; if it fails, \eg due to an invalid signature or other tampering, the vertex is ignored (see~\F\ref{fig:edg:corrupt_vertex}).

\sys adds edges to the EDG by iterating over vertices.\footnote{In practice, the FL job structure can be used to ascertain the vertices that \emph{should} be connected, thus avoiding naive iteration.} If the inputs of an EDR vertex match the outputs of another, an edge is added between the vertices, labeled with the dataflow between the exclaves~(lines~\ref{algo:edg-generate:edge-add-start}--\ref{algo:edg-generate:edge-add-end}). If the data was modified outside the exclave, \eg by an adversary during transit, there is no match, and no edge is added due to the input/output mismatch~(see~\F\ref{fig:edg:corrupt_edge}).

\begin{algorithm}[t]
\footnotesize
\caption{Constructing an EDG for auditing}
\label{algo:edg-generate}
\begin{algorithmic}[1] 
\setlength{\itemsep}{1pt} 
\setlength{\parsep}{0pt} 
\setlength{\topsep}{0pt} 
\REQUIRE Collection of vertices $v \in V'$, attestation reports $\sigma(v) \in \Sigma$
\ENSURE $EDG = (V, E, l)$
\STATE $V \gets \{v \mid \mathit{verify\_attestation\_report}(\sigma(v)) = \mathit{true}\}$ \label{algo:edg-generate:verify}
\STATE $EDG \gets \{V, E = \emptyset, l = \emptyset\}$
\STATE \textit{// Step 1: Index vertices by input}
\STATE \textbf{for each} $\{v_y \in V\}$ \textbf{do}
\STATE \hspace{1em} \textbf{for each} $\{input \in v_y.\mathit{in}\}$ \textbf{do}
\STATE \hspace{2em} \textbf{append} $v_y$ \textbf{to} $vertexOf[input]$ 
\STATE \textit{// Step 2: Create edge when input and output match}
\STATE \textbf{for each} $\{v_x \in V\}$ \textbf{do}
\STATE \hspace{1em} \textbf{for each} $\{output \in v_x.\mathit{out}\}$
\STATE \hspace{2em} \textbf{if} $\{vertexOf[output] \neq \emptyset \}$ \label{algo:edg-generate:edge-add-start}
\STATE \hspace{3em} \textbf{for each} $\{v_y \in vertexOf[output]\}$
\STATE \hspace{4em} $e \gets (v_y,v_x)$
\STATE \hspace{4em} $l(e) \gets v_y.\mathit{in} \cap v_x.\mathit{out} $ 
\STATE \hspace{4em} $E \gets E \cup \{e\}$, $l \gets l \cup l(e)$ \label{algo:edg-generate:edge-add-end}
\STATE \textbf{return} $(V, E, l)$ 
\end{algorithmic}
\end{algorithm}

\subsection{Checking verifiable claims}
\label{sec:flex:auditing}

\begin{figure*}[t]
  \centering
  \begin{subfigure}[t]{0.19\linewidth}
    \centering
    \includegraphics[width=\linewidth,keepaspectratio]{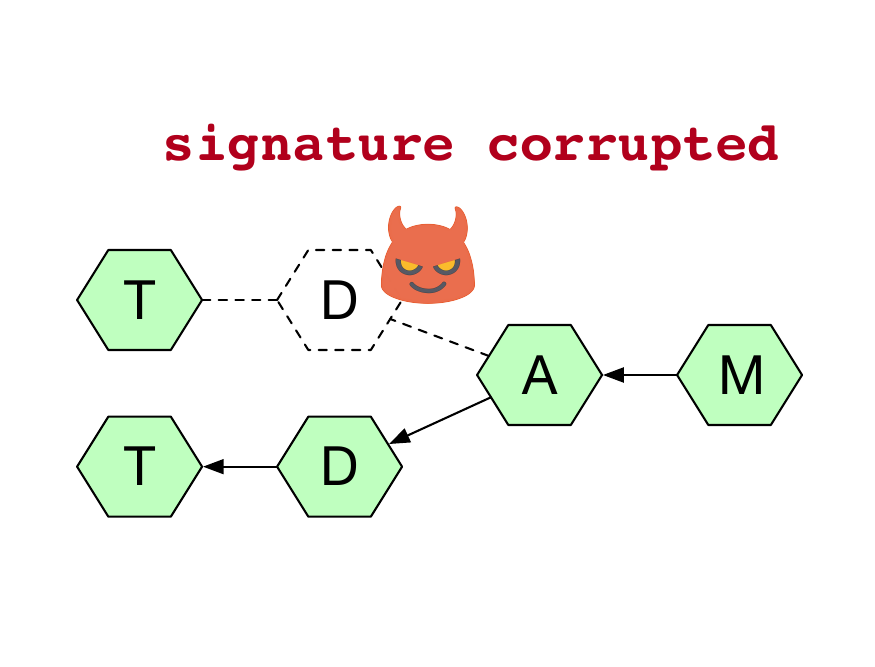}
    \caption{\normalfont Invalid attestation}
    \label{fig:edg:corrupt_vertex}
  \end{subfigure}
  \hfill
  \begin{subfigure}[t]{0.19\linewidth}
    \centering
    \includegraphics[width=\linewidth,keepaspectratio]{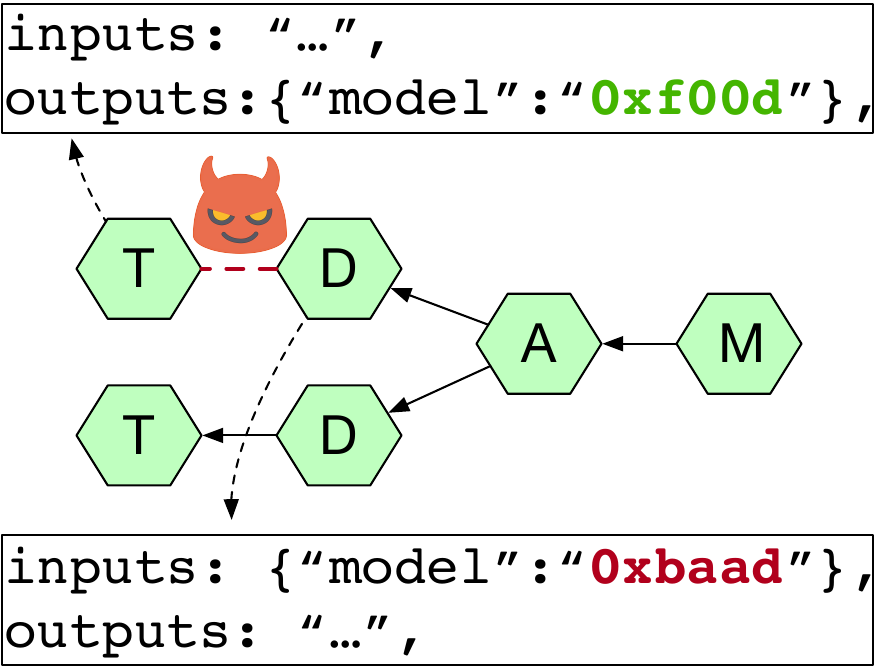}
    \caption{\normalfont Modifying data \\ in transit}
    \label{fig:edg:corrupt_edge}
  \end{subfigure}
  \hfill
  \begin{subfigure}[t]{0.19\linewidth}
    \centering
    \includegraphics[width=\linewidth,keepaspectratio]{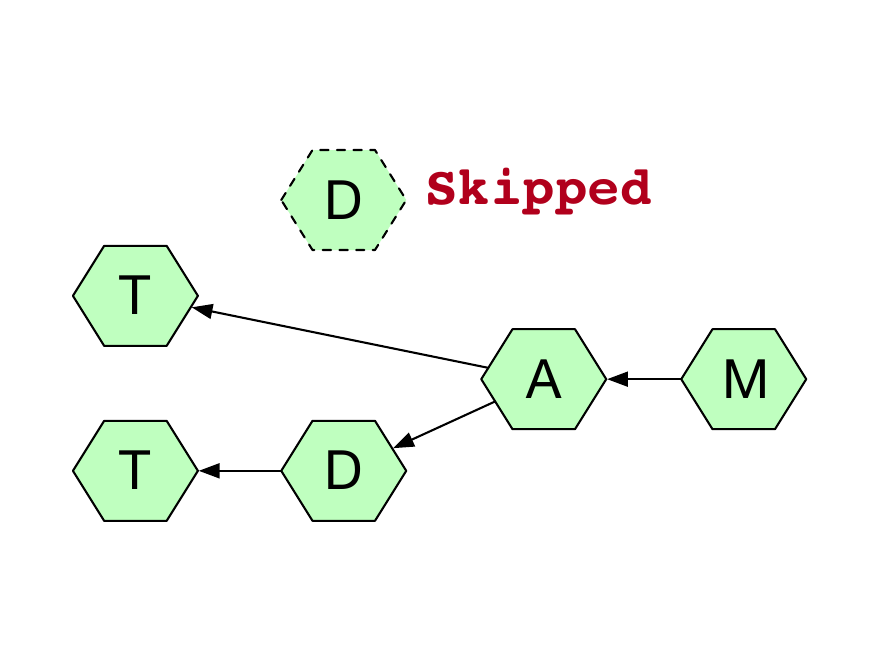}
    \caption{\normalfont Incorrect orchestration}
    \label{fig:edg:skip_vertex}
  \end{subfigure}
  \hfill
  \begin{subfigure}[t]{0.19\linewidth}
    \centering
    \includegraphics[width=\linewidth,keepaspectratio]{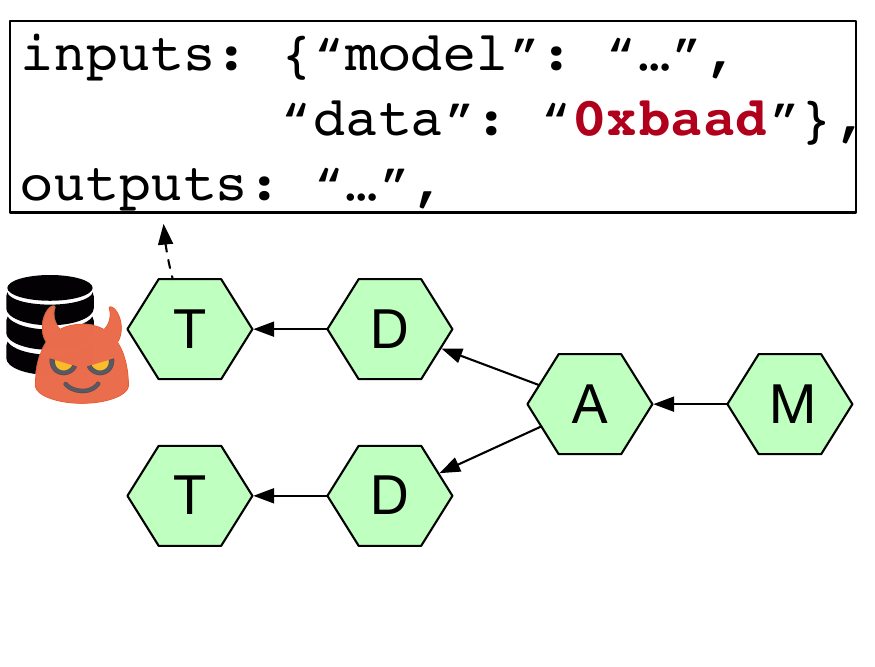}
    \caption{\normalfont Modified dataset}
    \label{fig:edg:runtime_data_poison}
  \end{subfigure}
  \hfill
  \begin{subfigure}[t]{0.19\linewidth}
    \centering
    \includegraphics[width=\linewidth,keepaspectratio]{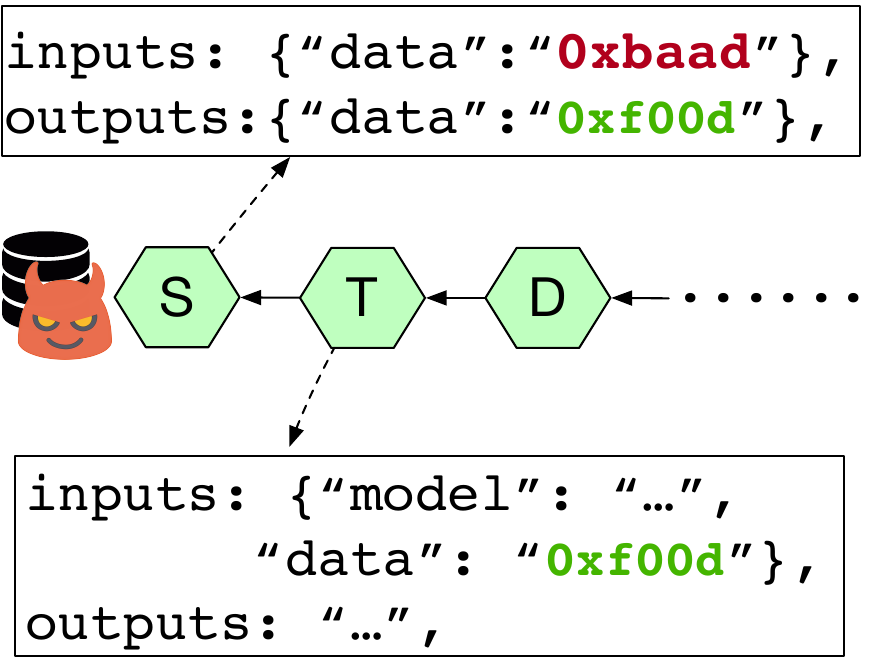}
    \caption{\normalfont Non-compliant dataset \\ with \underline{S}anitization}
    \label{fig:edg:data_sanitization}
  \end{subfigure}

  \caption{Variations of the EDG from \F\ref{fig:correct_edg} with deviations}
  \label{fig:edg_attack}
\end{figure*}

The constructed EDG is audited to verify claims about the FL model. Next, we describe the verifiable claims using examples of training-time deviations and explain how an auditor would detect them:

\mypar{Claim~1: Correct FL training algorithm implementation} Any participant could modify the code of tasks before deployment. For example, a data provider could modify the training and DP tasks to skip data integrity checks, ignore certain training data, inject too much noise, or modify model updates. A model provider could modify the aggregation or model update exclaves to omit inputs, change aggregation, modify model weights, or distribute different models to participants. The modified code would be measured by exclaves and recorded in the EDR. Since the auditor traverses the EDG and checks each vertex against a list of acceptable code measurements for tasks in the FL job, they would detect the claim violation.

\mypar{Claim~2: Correct transmission of training data} A malicious participant could modify the data in transit, either between exclaves~(\F\ref{fig:edg:corrupt_edge}), or when sent by the FL framework between participants. This would result in a mismatch between EDR outputs and inputs, omitting the edge from the EDG. Due to the missing output edge from a vertex, the auditor would identify the claim violation.

\mypar{Claim~3: Correct use of DP/unbiased data aggregation} As shown in \F\ref{fig:edg:skip_vertex}, a malicious participant could bypass the DP task execution and send the training task's output directly without DP noise to the model provider, \eg to increase the impact of their harmful data records. Similarly, a model provider could introduce bias by dropping contributions from certain data providers. This is more subtle than the previous attacks, because it still results in a complete EDG. The shape of the resulting EDG, however, would differ: it would miss vertices/edges, or have edges that span multiple training rounds, which would be detectable by the auditor.

\mypar{Claim~4: Integrity of training dataset} A malicious data provider could add, remove, or modify a dataset after training has commenced (\eg performing an adaptive poisoning attack~\cite{a3fl}). If the data is modified while being read by an exclave, the verification against the Merkle tree fails; if the dataset is modified between training rounds~(see~\F\ref{fig:edg:runtime_data_poison}), the hash of the modified dataset is recorded in the EDG, allowing the auditor to discover the change.

\mypar{Claim~5: Appropriate dataset sanitization} A data provider may use a non-compliant dataset without sanitization, \eg with harmful content or personal information. If the FL job includes a sanitization task, its correct execution is recorded by the EDRs. Therefore, an auditor can compare the training task's input dataset commitment against the set of data commitments produced by the sanitization task~(\F\ref{fig:edg:data_sanitization}). An adversary may skip sanitization, or excessively filter their data to contribute less; both are detectable due to the code measurement of the data sanitization task in the EDG. Note that the auditor does not require access to the plaintext dataset to detect whether the dataset was exposed to a given sanitization task: EDGs only contain the data hashes, because \sys audits the training process rather than the training data.



\section{\sys Implementation}
\label{sec:impl}

\begin{figure}[t]
  \centering
  \includegraphics[width=0.48\textwidth]{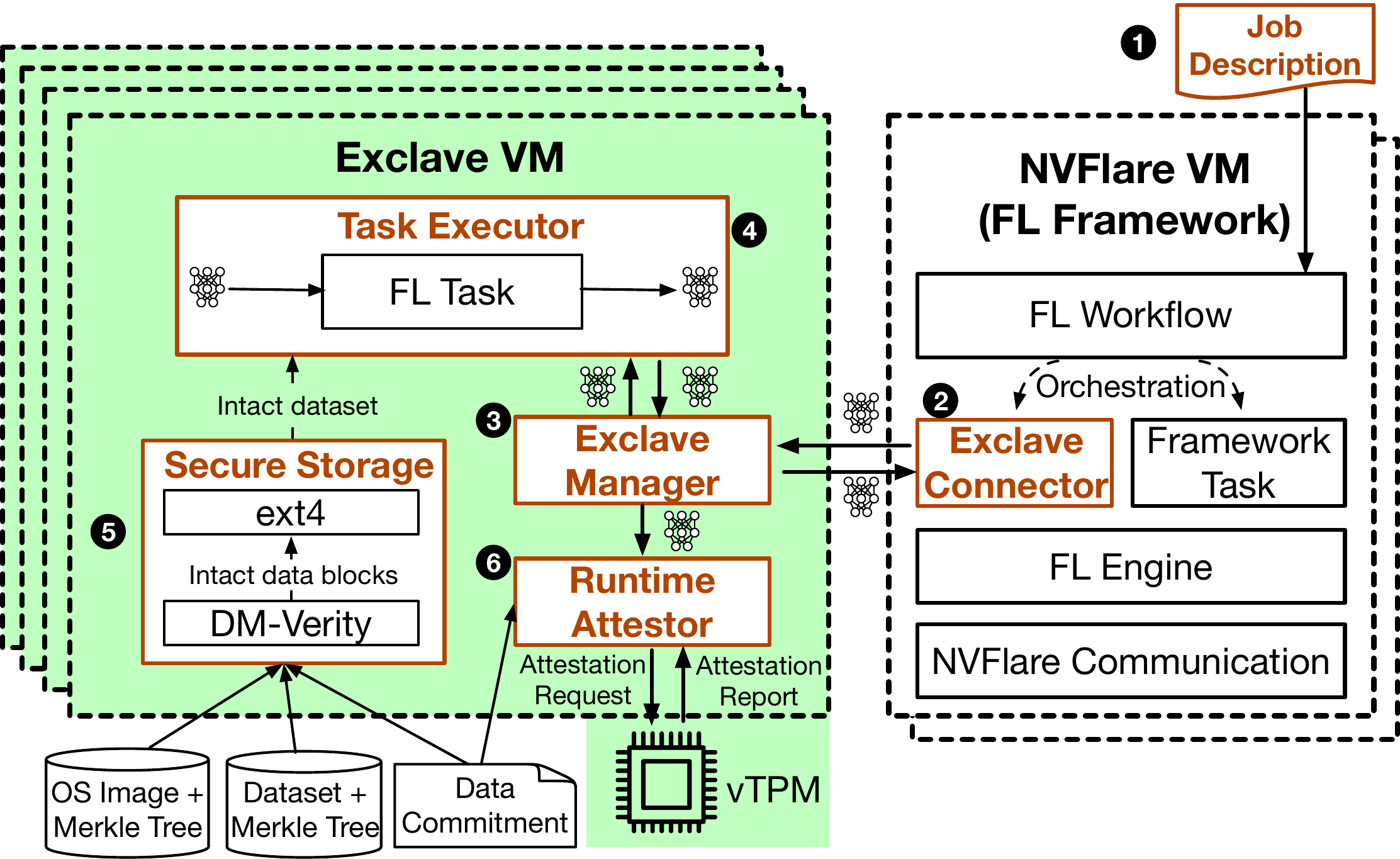}
  \caption{\sys implementation}
  \label{fig:arch}
\end{figure}

In this section, we describe the \sys implementation~(\S\ref{sec:impl:overview}) and its components~(\S\ref{sec:impl:components}), as well as our implementation of exclave attestation~(\S\ref{sec:impl:attestation}) and integrity-protected data storage~(\S\ref{sec:impl:storage}).

\subsection{Overview}
\label{sec:impl:overview}

A goal of the \sys implementation is to be compatible with different FL frameworks and to support existing FL frameworks with minimal code changes. To achieve this, \sys{} adopts an orchestrator/worker model in which an untrusted FL framework orchestrates the FL job and task execution inside exclaves.

\F\ref{fig:arch} shows how \sys integrates with an FL Framework: a model provider submits an FL job description to the FL Framework~\aptLtoX{\ding{182}}{\myc{1}}. The framework orchestrates the job according to its description, and allocates FL tasks to exclaves via Exclave Connectors~\aptLtoX{\ding{183}}{\myc{2}}, which are \sys{} components that invoke exclaves. When an exclave receives a task, an Exclave Manager~\aptLtoX{\ding{184}}{\myc{3}} forwards its payload (\eg global model) to a Task Executor. During task execution~\aptLtoX{\ding{185}}{\myc{4}}, the task can access persistent storage via a Secure Storage module~\aptLtoX{\ding{186}}{\myc{5}}, which provides integrity protection for on-disk datasets. When the task completes, a Runtime Attestor~\aptLtoX{\ding{187}}{\myc{6}} attests the EDR of the exclave. The execution result (\eg a local model) is then returned.

\begin{table}[t]
    \centering
    \caption{FL tasks deployed in exclaves}\label{tab:api}
    \small
    \begin{tabular}{lll}
    \toprule
    \textbf{Exclave}                                             & \textbf{Task API}                                                                   & \textbf{Description}                   \\
    \midrule
    Training                                    & \texttt{StartTrain(...)}                                                       & Receives global model          \\
					 &		\texttt{GetWeight(...)}                                                        & Returns local model \\
    \midrule
    DP                                        & \texttt{AddNoise(...)}                                                         & Performs local DP              \\
    \midrule
    Aggregation                                 & \texttt{Accept(...)}                                                           & Receives local models          \\
					 & \texttt{Aggregate(...)}                                                        & Aggregates local models        \\
    \midrule
    Model Update                                & \texttt{ModelUpdate(...)}                                                      & Applies diff to full model \\
    \bottomrule
    \end{tabular}
\end{table}

\subsection{Components}
\label{sec:impl:components}

Our \sys implementation has the following components:

\mypar{FL Framework} \sys uses NVFlare~\cite{roth2022nvidia}, which is an open-source distributed FL framework by Nvidia. It provides FL task implementations for many FL algorithms, \eg federated averaging, federated optimization, and secure aggregation, and other building blocks, \eg model trainers, model aggregators, and task filters. \T\ref{tab:api} shows a set of FL tasks used in a typical FL job. \sys executes all of these tasks inside exclaves.

\myparr{Exclave Connectors} are gRPC clients~\cite{grpc} that NVFlare uses to execute FL tasks in exclaves. They are implemented as NVFlare components, which makes them compatible with the NVFlare API. To add an exclave, \sys includes an Exclave Connector in NVFlare's job description file. This approach of using Exclave Connectors avoids direct modifications to the NVFlare code.

\myparr{Exclaves} are implemented as confidential VMs~(CVMs) using AMD SEV-SNP~\cite{sev2020strengthening}. Within each CVM, gRPC servers can execute FL tasks. 

\myparr{Exclave Managers} expose exclave interfaces to NVFlare to support the orchestration of FL jobs. Exclave Managers receive FL tasks from NVFlare, deserialize payloads, and forward payloads to the Task Executor. They then take the execution results from the Task Executor and pass them to NVFlare.

\myparr{Task Executors} execute tasks inside exclaves. They receive FL tasks from the Exclave Manager, load the data/code, and execute the tasks. For example, the Task Executor for LLM training loads the model, the training code, and the training dataset. \sys reuses FL task execution code from NVFlare with minimal changes.

\tinyskip

\noindent
The \textbf{Runtime Attestor} constructs EDRs and interacts with a virtual Trusted Platform Module~(vTPM)~\cite{vtpm} for attestation. During task execution, it records the storage accesses and monitors requests/responses passing through the Exclave Manager to get exclaves' input and output data. It stores a hash of each data item together with metadata describing the intended usage (\eg input/output) and properties (\eg model/dataset). After an input/output pair is collected, it builds a nested JSON object: keys contain the data usage and properties, and the values contain data hashes. The Runtime Attestor invokes the vTPM's signing API with a SHA-256 digest of this object, which the vTPM signs together with a system measurement to produce an EDR.

\subsection{Attestation} 
\label{sec:impl:attestation}

To prototype exclave attestation without firmware modifications, we use a vTPM~\cite{vtpm} and virtual machine privilege levels~(VMPL)~\cite{sev2020strengthening}. We run the vTPM in VMPL~0 (highest privilege), and exclaves in VMPL~2 (lower privilege); VMPL~1 remains unused, following existing AMD SEV-SNP deployment practices~\cite{svsm}. The MMU and RMP enforce isolation: lower privilege levels (\ie exclaves) cannot interfere with higher ones (\ie vTPM), protecting the attestation process. The vTPM provides attestation primitives (measurement and signing), and its signing key is endorsed by the PSP on CVM launch. Attestation requests/responses pass through the RMP-protected, VMPL~2 memory region, writable by the vTPM and exclaves. The vTPM reads the runtime claim's hash, extends it into PCR23, signs the PCRs, and writes the signed report.

This approach makes the vTPM an emulation of AMD's PSP with our proposed firmware modifications. While this provides a significant degree of security, there is residual risk because the vTPM runs on the main CPU and shares micro-architectural state (\eg CPU caches) with attacker-controlled processes. While having the vTPM at a higher privileged VMPL increases attack difficulty, moving the attestation signing key to a separate co-processor, as done by exclaves, would eliminate this risk. This also aligns with the planned evolution of TEEs: \eg in future generations, Intel~TDX~\cite{cheng2023intel} proposes to create  attestation quotes using separate security hardware~\cite{tdxquote}.

\subsection{Secure storage with commitment}
\label{sec:impl:storage}

The Secure Storage module offers integrity-preserving, verifiable storage for exclaves, which protects both datasets and task code.

\mypar{Datasets} \sys uses Linux's \emph{dm-verity}~\cite{dmverity} for integrity protection, which checks accessed disk blocks against a Merkle tree. Data providers use the dm-verity tool to package datasets with a Merkle tree. The generated data commitment is a unique measurement over the dataset, which is included in EDRs by the Runtime Attestor. When setting up an exclave, \sys mounts the data disk and checks the data commitment against the Merkle tree.

\mypar{Code} \sys also uses dm-verity protection for the exclave's code image by using the Merkle tree root hash as a code measurement. When an exclave is launched, the vTPM includes the code measurement in the \texttt{PCR11} field of the attestation report. Since it uniquely identifies the exclave code, it can be used by the auditor to determine if an FL task was executed by the correct exclave.



\begin{figure*}[t] 
    \centering
    \begin{subfigure}[b]{0.32\textwidth}
        \includegraphics[trim={0 0.6cm 0 0}, clip, width=\textwidth]{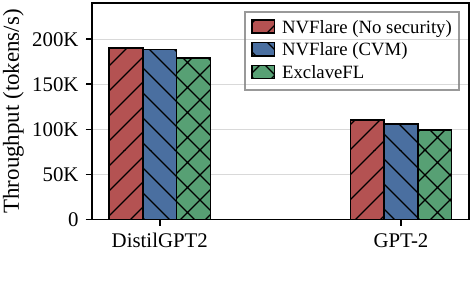}
        \caption{\textnormal{Language modeling}}
        \label{fig:wiki_e2e}
    \end{subfigure}
    \hfill
    \begin{subfigure}[b]{0.32\textwidth}
        \includegraphics[trim={0 0.6cm 0 0}, clip, width=\textwidth]{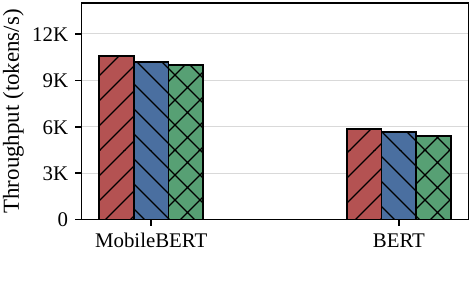}
        \caption{\textnormal{Named-entity recognition}}
        \label{fig:ncbi_e2e}
    \end{subfigure}
    \hfill    
    \begin{subfigure}[b]{0.32\textwidth}
        \includegraphics[trim={0 0.6cm 0 0}, clip, width=\textwidth]{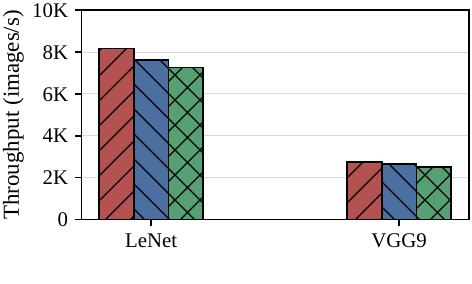}
        \caption{\textnormal{Image classification}}
        \label{fig:cifar10_e2e}
    \end{subfigure}
    \caption{End-to-end training throughput}
    \label{fig:e2e}
\end{figure*}

\section{Evaluation}
\label{sec:eval}

We evaluate \sys to answer: (i)~what is \sys{}'s end-to-end training performance compared to existing solutions?~(\hyperref[sec:eval:e2e]{§\ref{sec:eval:e2e}}); (ii)~how does \sys scale with more data providers?~(\hyperref[sec:eval:scalability]{§\ref{sec:eval:scalability}}); (iii)~what is the latency breakdown of \sys?~(\hyperref[sec:eval:breakdown]{§\ref{sec:eval:breakdown}}); (iv)~what overhead does \sys{}'s runtime attestation add?~(\hyperref[sec:eval:attestation]{§\ref{sec:eval:attestation}}); (v)~what is the impact of secure storage on throughput and latency?~(\hyperref[sec:eval:storage]{§\ref{sec:eval:storage}}); and (vi)~what is the auditing overhead as \sys scales?~(\hyperref[sec:eval:auditor]{§\ref{sec:eval:auditor}})

\subsection{Experimental setup}
\label{sec:eval:setup}

\mypar{Testbed} We deploy confidential VMs~(CVMs) on Azure: CVMs without GPUs run on AMD EPYC 7763v CPUs; CVMs with GPUs run on AMD EPYC 9V84 CPUs, both with SEV-SNP and vTPM enabled. The GPUs are NVIDIA~H100 with Hopper Confidential Computing extensions~\cite{h100cc}. We deploy 5~CVMs (EC4as\_v5) to execute 1~model update exclave and 4~NVFlare clients; 5~CVMs (EC8as\_v5) to execute 4~DP exclaves and 1~NVFlare server; and 1~CVM (EC16as\_v5) as the aggregation exclave. Training exclaves are deployed on 4~CVMs with CPUs (DC32as\_v5) or 4~CVMs with GPUs (NCC40ads\_H100\_v5). Each training exclave uses an Azure P30 data disk (200\unit{MB/s} of throughput with 5k\unit{IOPS}). 

\mypar{Implementation} We implement \sys in Python and C++. It uses PyTorch~v2.5.0 with CUDA~12.4 for the FL tasks, Azure Guest Attestation Library~v1.0.2, and NVFlare~v2.2.1~\cite{roth2022nvidia}. As described in \S\ref{sec:impl:attestation}, we emulate the attestation behavior of exclaves using a vTPM implementation~\cite{vtpm} running on the AMD SEV-SNP platform.

\mypar{Baselines} We compare \sys with two baselines: (i)~\NVFlareNoSec executes each NVFlare node (server and clients) in its own regular VM without SEV-SNP features; and (ii)~\NVFlareCVM~\cite{guo2024trustworthy} executes each NVFlare node in its own CVM. To facilitate comparison, both \NVFlareCVM and \sys use the vTPM for attestation and dm-verity for secure storage.

\mypar{Datasets} We use 3~public datasets: (i)~Wikitext103~\cite{merity2016pointer} consists of raw text with over 100\unit{M} tokens from Wikipedia articles; (ii)~NCBI-disease~\cite{dougan2014ncbi} is a disease text mining dataset with 793~fully annotated PubMed abstracts; and (iii)~CIFAR10~\cite{krizhevsky2009learning} has annotated $32\times32$ pixel images with 50,000~training and 10,000~testing samples.

\mypar{Models} We use 6~widely-used models, covering three different categories: (i)~decoder-only transformers---we use GPT-2~\cite{radford2019language} with 124\unit{M} parameters, and a distilled version of GPT-2, DistilGPT2~\cite{sanh2019distilbert}, with 82\unit{M} parameters; (ii)~encoder-only transformers---we employ BERT~\cite{devlin2019bert} with 110\unit{M} parameters, and MobileBERT~\cite{sun2020mobilebert} with 25\unit{M} parameters; and (iii)~convolutional neural networks~(CNN)---we use LeNet~\cite{lecun1998gradient}, a 5-layer CNN, and VGG9~\cite{simonyan2014very}, a 9-layer CNN.

\begin{figure*}[t] 
    \centering
    \begin{subfigure}[b]{0.32\textwidth}
    \includegraphics[width=\textwidth]{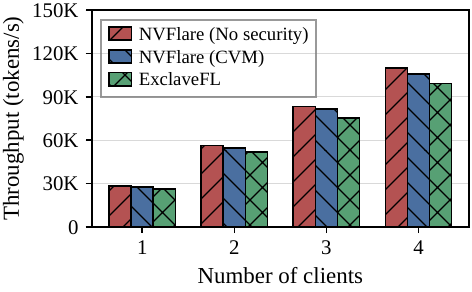}
        \caption{\textnormal{GPT-2}}
        \label{fig:gpt2_scalability}
    \end{subfigure}
    \hfill
    \begin{subfigure}[b]{0.32\textwidth}
	\includegraphics[width=\textwidth]{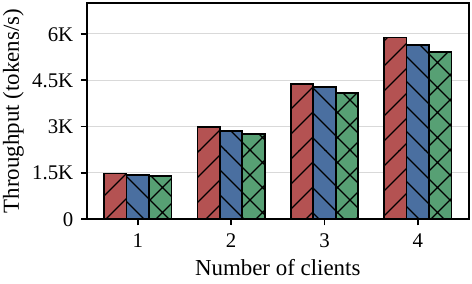}
	\caption{\textnormal{BERT}}
        \label{fig:bert_scalability}
    \end{subfigure}
    \hfill    
    \begin{subfigure}[b]{0.32\textwidth}
	\includegraphics[width=\textwidth]{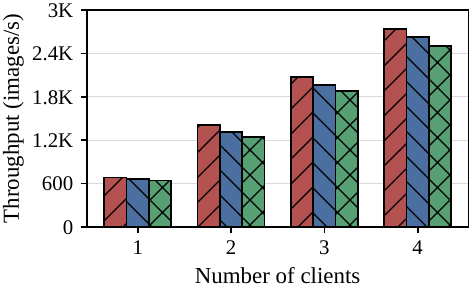}
	\caption{\textnormal{VGG9}}
        \label{fig:vgg9_scalability}
    \end{subfigure}
    \caption{Throughput with different numbers of clients}
    \label{fig:scalability}
\end{figure*}

\subsection{End-to-end throughput}
\label{sec:eval:e2e}

We evaluate the end-to-end performance of \sys in terms of training throughput. We run \sys with 1~FL server and 4~FL clients. Each client performs model training with either a CPU or GPU CVM, depending on the workload. After each training round, the FL clients use the sparse vector differential privacy technique~(SVT-DP)~\cite{lyu2016understanding} to add DP noise to the locally trained model. \sys runs each FL job for 10~rounds with federated averaging~(FedAvg) aggregation. We measure \sys{}'s performance for our three ML categories: (i)~language modeling; (ii)~named-entity recognition; and (iii)~image classification. \F\ref{fig:e2e} shows the measured throughput results.

\mypar{Language modeling} In this experiment, each client has a training exclave that executes the language modeling task in a GPU CVM (40~vCPUs; 320\unit{GB} of RAM; 1~NVIDIA H100 GPU). We train both GPT-2 and DistilGPT2 on the Wikitext103 dataset, which is partitioned into four subsets, one for each client.

\F\ref{fig:wiki_e2e} shows the training throughput. With DistilGPT2, \sys has a $6\%$ overhead compared to the \NVFlareNoSec baseline; \NVFlareCVM has a $1\%$ overhead. With GPT-2, when compared to \NVFlareNoSec, \sys and \NVFlareCVM incur less than $10\%$ and around $4\%$ overhead in training throughput, respectively. We observe that, in both cases, \sys is slower than \NVFlareCVM: in \sys, exclaves introduce runtime overhead due to the runtime attestation; and the distributed deployment of exclaves adds network overhead when (de)serializing the model and transmitting the serialized model.

\mypar{Named-entity recognition} Training exclaves execute in CPU CVMs (32~vCPUs; 128\unit{GB} of RAM). We use the MobileBERT and BERT models on the NCBI-disease dataset, split into 4~client shards.

\F\ref{fig:ncbi_e2e} shows the training throughput. With MobileBERT, \sys and \NVFlareCVM are $5\%$ and $4\%$ slower than \NVFlareNoSec, respectively. With BERT, \sys incurs an $8\%$ overhead compared to \NVFlareNoSec, and \NVFlareCVM has a $4\%$ performance cost. We conclude that, when training MobileBERT, \sys has less performance impact compared to the BERT training: since MobileBERT is $4.4\times$ smaller than BERT, there is less network communication between exclaves in \sys.

\mypar{Image classification} We also execute the image classification task with CPU CVMs under the same setup as before. Here, we use the LeNet and VGG9 models with the CIFAR10 dataset. Since the CIFAR10 dataset is small, we replicate it four times before sharding, ensuring that each client has sufficient training data.

\F\ref{fig:cifar10_e2e} shows the performance results. For VGG9 training, \sys and \NVFlareCVM compared to \NVFlareNoSec incur $9\%$ and $4\%$ overhead, respectively. Compared to other workloads, \sys has a higher overhead for LeNet training: it is $11\%$ slower than \NVFlareNoSec. Even with \NVFlareCVM, LeNet training overhead is 7\%. Here, the CVM overhead and \sys's extra scheduling/communication impact performance.

For moderate and large models~(\ie VGG9, MobileBERT, BERT, DistilGPT2, and GPT-2), \sys{}'s overhead is below $10\%$ compared to \NVFlareNoSec, which is slightly higher than \NVFlareCVM{}'s overhead of up to $4\%$. While LeNet shows higher overhead than other models, the overhead stays low across all models.

\aptLtoX{\begin{figure}[t]
  \centering
    \centering
    \includegraphics[width=.32\linewidth]{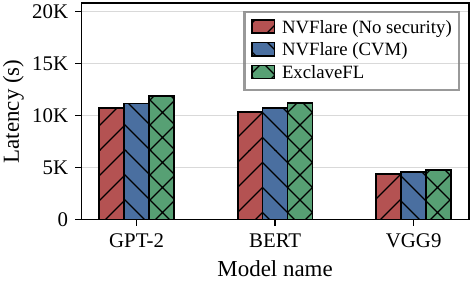}
    \caption{End-to-end FL latency}
    \label{fig:latency}
  \end{figure}
  \begin{figure}
    \centering
    \includegraphics[width=.32\linewidth]{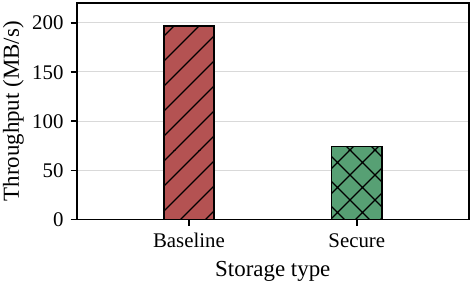}
    \caption{Storage read throughput}
    \label{fig:storage-throughput}
  \end{figure}
  \begin{figure}
    \centering
    \includegraphics[width=.32\linewidth]{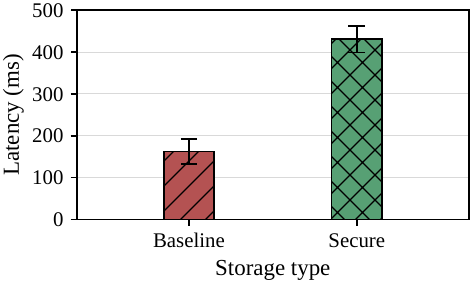}
    \caption{Storage read latency}
    \label{fig:storage-latency}
\end{figure}}{
\begin{figure*}[t]
  \centering
  \begin{minipage}[t]{0.32\textwidth}
    \centering
    \includegraphics[width=\linewidth]{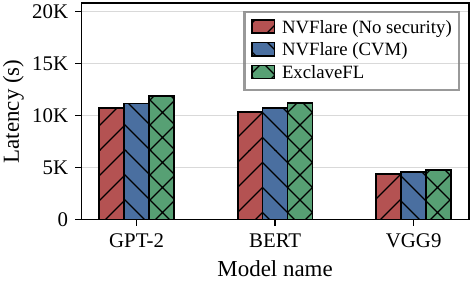}
    \caption{End-to-end FL latency}
    \label{fig:latency}
  \end{minipage}\hfill
  \begin{minipage}[t]{0.32\textwidth}
    \centering
    \includegraphics[width=\linewidth]{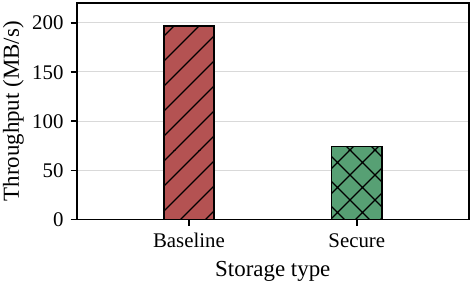}
    \caption{Storage read throughput}
    \label{fig:storage-throughput}
  \end{minipage}\hfill
  \begin{minipage}[t]{0.32\textwidth}
    \centering
    \includegraphics[width=\linewidth]{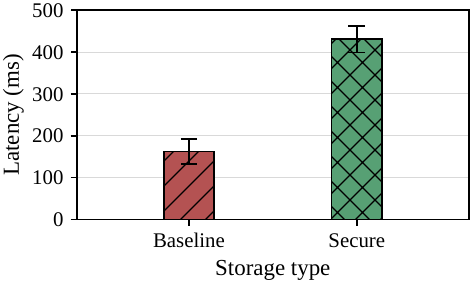}
    \caption{Storage read latency}
    \label{fig:storage-latency}
  \end{minipage}
\end{figure*}}

\subsection{Scalability}
\label{sec:eval:scalability}

Next, we investigate \sys{}'s scalability. We increase the number of FL clients and measure training throughput. Note that adding FL clients results in a corresponding increase in data parallelism. We use the same testbed settings and FL tasks as in \hyperref[sec:eval:e2e]{§\ref{sec:eval:e2e}}.

First, we measure \sys{}'s scalability with the GPT-2 model for language modeling (\F\ref{fig:gpt2_scalability}). Here, we scale \sys from 1 to 4~clients, each with a GPU. With $4\times$ the number of FL clients, the throughput of \sys, \NVFlareCVM, and \NVFlareNoSec increases to $3.8\times$, $3.8\times$, and $3.9\times$, respectively.

We also consider the BERT~(\F\ref{fig:bert_scalability}) and VGG9~(\F\ref{fig:vgg9_scalability}) models. In these experiments, we increase the CPU count when scaling FL clients. With BERT, scaling the clients $4\times$ leads to $3.9\times$, $4.0\times$, and $4.0\times$ the throughput for \sys, \NVFlareCVM, and \NVFlareNoSec, respectively. For VGG9, we observe that the throughput for \sys, \NVFlareCVM, and \NVFlareNoSec increases to $3.9\times$, $4.0\times$, and $4.0\times$, respectively, when we quadruple the client count. Both results show that \sys scales effectively with more FL clients, similar to the baselines. Since all clients run in parallel, runtime and I/O overheads do not accumulate.

\begin{table}[t]
\centering
\caption{Latency for unmodified NVFlare (No~Sec) and \sys}
\small
\setlength{\tabcolsep}{3pt}
\begin{tabular}{lrrrrrr}
\toprule
 & \multicolumn{2}{c}{\textbf{GPT-2}} & \multicolumn{2}{c}{\textbf{BERT}} & \multicolumn{2}{c}{\textbf{VGG9}} \\
\cmidrule(lr){2-3}\cmidrule(lr){4-5}\cmidrule(lr){6-7}
\textbf{Component} & \textbf{No~Sec} & \textbf{ExFL} & \textbf{No~Sec} & \textbf{ExFL} & \textbf{No~Sec} & \textbf{ExFL} \\
\midrule
Training      & 8,387\unit{s}   & 8,659\unit{s}   & 8,745\unit{s}   & 9,164\unit{s}   & 4,195\unit{s}   & 4,641\unit{s}  \\
DP            & 1,035\unit{s}   & 1,176\unit{s}   & 699\unit{s}     & 763\unit{s}     & 20\unit{s}      & 21\unit{s}     \\
Aggregation   & 9\unit{s}       & 50\unit{s}      & 6\unit{s}       & 29\unit{s}      & $<$1\unit{s}    & 1\unit{s}      \\
Model update  & 5\unit{s}       & 29\unit{s}      & 2\unit{s}       & 18\unit{s}      & $<$1\unit{s}    & 1\unit{s}      \\
FL framework  & 1,286\unit{s}   & 1,983\unit{s}   & 883\unit{s}     & 1,257\unit{s}   & 158\unit{s}     & 124\unit{s}    \\
\midrule
Total         & 10,721\unit{s}  & 11,897\unit{s}  & 10,335\unit{s}  & 11,232\unit{s}  & 4,373\unit{s}   & 4,789\unit{s}  \\
\bottomrule
\end{tabular}
\label{tab:breakdown}
\end{table}

\subsection{Latency breakdown}
\label{sec:eval:breakdown}

We report \sys{}'s end-to-end latency in~\F\ref{fig:latency}. While latency is inversely related to throughput, it makes the system overhead explicit and can be decomposed into per-component contributions. We provide such a latency breakdown for \NVFlareNoSec and \sys in \T\ref{tab:breakdown}. Since exclaves use dm-verity-protected secure storage, the storage overhead is included in both the end-to-end latency results~(\F\ref{fig:latency}) and the latency breakdown~(\T\ref{tab:breakdown}).

Among all components, the training exclave dominates execution time (97\% for VGG9; 82\% for BERT; 73\% for GPT-2). Note that GPT-2 trains with GPU exclaves, while VGG9 and BERT train with CPU exclaves. Thus, the GPT-2 training task occupies a smaller proportion of overall time.

The time spent executing the aggregation and model update exclaves increases linearly with model size, because the averaging/addition operations have linear time complexity. The DP exclave has a worst-case quadratic time complexity, because the SVT-DP algorithm iterates over all model parameters (potentially multiple times) to calculate the noise and apply the noise threshold. The total time spent on the DP, aggregation, and model update exclaves is less than 1\% for VGG9 and up to 11\% for GPT-2.

The execution time for the FL framework, including the time spent on orchestration and communication, also increases linearly with model size. Frequent communication and orchestration make the FL framework the second largest contributor to \sys{}'s execution time: 17\%, 11\%, and 3\% for GPT-2, BERT, and VGG9, respectively. This overhead comes from \sys{}'s orchestration of tasks that execute in exclaves: it is necessary for \sys to deploy exclaves in separate CVMs to provide the fine-grained runtime attestation of each exclave.

Compared to \NVFlareNoSec, \sys{}'s overhead comes from two sources. First, running FL workloads across disaggregated exclaves adds extra communication and orchestration overheads that inflate the FL framework cost. These overheads are significant for BERT and GPT-2, because these models are large. For VGG9, the communication overhead is small.\footnote{The residual difference is due to cloud placement noise: for \NVFlareNoSec, the FL framework spends more time waiting for straggling clients.} Second, each exclave incurs runtime costs from running the workload inside a CVM with a gRPC server, together with model (de)serialization/hashing, secure storage, and attestation. This appears as the per-component overhead of \sys over \NVFlareNoSec in~\T\ref{tab:breakdown}.

In addition, we evaluate the overhead of \sys's data sanitization exclave, which a data provider runs prior to the FL training on an Azure NCC40ads\_H100\_v5 CVM. The data sanitization exclave uses a classification model for harmful text, granite-guardian-hap-38m~\cite{granite}, to detect toxic or abusive content and then removes the identified content to sanitize the dataset. 

We measure the execution time of the data sanitization exclave when sanitizing each data provider's Wikitext103 dataset. The result shows that the exclave takes 257~seconds to sanitize each data provider's dataset, which adds 2\% overhead to \sys's GPT-2 training. We conclude that enforcing dataset properties using standard sanitization techniques is therefore feasible in practice.

\subsection{Overhead of runtime attestation}
\label{sec:eval:attestation}

Next, we measure the overhead due to runtime attestation. We use a single CVM with 4~vCPUs and 16\unit{GB} of RAM to generate a SHA-256 digest over 1\unit{GB} of input data and include the resulting hash digest as the runtime claim in an attestation request.

We observe that, on average, creating and signing an attestation report with the vTPM takes 4\unit{ms}. This step has constant time complexity, because both creation and signing operations take a fixed-size hash digest as input. Generating hash digests is the most time-consuming operation during attestation: it takes 687\unit{ms} to calculate a SHA-256 hash over the 1\unit{GB} of input data, and the time grows linearly with the input data size.

We also measure the attestation overhead with AMD's PSP: the CVM requires less than 7\unit{ms} to fetch a signed attestation report from the PSP. Although we use vTPM attestation to emulate PSP attestation, the performance difference between them is marginal (under 3\unit{ms} per attestation), making it negligible compared to \sys{}'s end-to-end training time~(hours).

\subsection{Overhead of secure storage}
\label{sec:eval:storage}

We also explore the performance impact of \sys{}'s secure storage for datasets. We mount a disk protected by dm-verity on an Azure~DC32as\_v5 CVM (P30 data disk with up to 200\unit{MB/s} and 5,000\unit{IOPS}). We then use the FIO benchmark~\cite{fio}, which sequentially reads data from the ext4 file system on the secure disk. We compare against insecure storage as a baseline, which uses a standard D32as\_v5 VM on an unprotected Azure~P30 disk.

As \F\ref{fig:storage-throughput} shows, the sequential read throughput of the integrity-protected disk is 62\% lower compared to regular storage. This is due to the cost of checking the block against the Merkle hash tree, together with CVM overhead. Latency also increases accordingly~(\F\ref{fig:storage-latency}): the secure storage module has $2.7\times$ the baseline latency (reported latencies include queuing time from the in-flight I/Os). The resulting read amplification impacts both throughput and latency, which contributes to \sys{}'s performance overhead.

\begin{table}[t]
\centering
\caption{Overhead of auditing}\label{tab:auditor}
\small
\begin{tabular}{lrrrr}
\toprule
  \textbf{\#Clients $\times$} & \textbf{Signature}   & \textbf{EDG} & \textbf{Total} & \textbf{Storage} \\
  \textbf{\#Rounds} & \textbf{Verification} & \textbf{Creation} &  &  \\
  
  \midrule
10 $\times$ 10    & 26\unit{ms}    &  3\unit{ms}    &  29\unit{ms}     &   <1\unit{MB}  \\ 
10 $\times$ 100   & 196\unit{ms}   &  13\unit{ms}   &  209\unit{ms}    &   2\unit{MB}   \\ 
10 $\times$ 1,000  & 1,886\unit{ms}  &  131\unit{ms}  &  2,017\unit{ms}   &   17\unit{MB}  \\ 
100 $\times$ 10   & 218\unit{ms}   &  20\unit{ms}   &  238\unit{ms}    &   2\unit{MB}   \\ 
100 $\times$ 100  & 1,802\unit{ms}  &  146\unit{ms}  &  1,948\unit{ms}   &  16\unit{MB}   \\ 
  100 $\times$ 1,000 & 17,404\unit{ms} &  1,837\unit{ms} &  19,241\unit{ms}  &  162\unit{MB}  \\ 
  \bottomrule
\end{tabular}
\end{table}

\subsection{Overhead of auditing}
\label{sec:eval:auditor}

We evaluate the overhead that the auditing process incurs in real-world settings, when the number of clients scales from 10 to 100, and the number of rounds varies from 10 to 1,000. We generate synthetic EDRs and measure the auditor's computation and storage overheads in a standard VM with 4~vCPUs and 16\unit{GB} of RAM.

As shown in \T\ref{tab:auditor}, when 100~clients train for 1,000~rounds, it takes the auditor 19~seconds to verify all EDR signatures and construct the EDG, which is negligible in comparison to the FL training time.\footnote{Verifying each exclave signing key against its endorsement is a one-time, per-exclave startup cost, and thus not included in the costs measured for each EDR.} The incurred storage cost during auditing is 162\unit{MB}. We conclude that \sys{}'s auditing process introduces a small overhead, even with a large number of clients and training rounds.



\section{Related Work}
\label{sec:rel_work}

\mypar{Mitigating attacks in FL} Aggregation protocols can be designed to mitigate privacy/integrity attacks in FL. \emph{Secure aggregation}~\cite{bonawitz2017secure, sharma2019secure} addresses FL privacy concerns when aggregators are untrusted through cryptographic means. For example, Bonawitz~\etal\cite{bonawitz2017secure} implement secure aggregation using pairwise Diffie-Hellman key agreement and t-out-of-n Shamir's secret sharing; Sharma~\etal\cite{sharma2019secure} use additive secret sharing. These protocols are complementary to \sys{}, because they ensure privacy, while \sys provides verifiable integrity guarantees that the correct computation was performed. Secure aggregation protocols could be used with \sys to provide both privacy and integrity guarantees. \sys is compatible with a range of FL-specific techniques: \emph{robust aggregation}~\cite{pillutla2022robust} makes the FL training process more resilient against integrity attacks; Tolpegin~\etal\cite{tolpegin2020data} detect data poisoning attacks using outlier detection. To protect FL training against model poisoning attacks, Krum~\cite{blanchard2017machine} and Trimmed-mean~\cite{yin2018byzantine} are \emph{Byzantine-robust aggregation} protocols. While such protocols require at least a majority of trusted data providers, \sys{}'s verifiable claims hold with any number of malicious participants.

\mypar{Trusted execution for FL} Mo~\etal\cite{mo_ppfl_2021} assume that both the FL server and clients are honest-but-curious. They use TEEs to prevent participants from observing the model during training, thus preventing inference or data reconstruction attacks. Messaoud~\etal\cite{valerio21gradsec} address client-side privacy threats by protecting gradient updates inside TEEs. Since a subset of the model layers is protected from gradient leakage, this mitigates inference attacks. Both of the approaches assume a weaker threat model, without \sys{}'s detection of integrity attacks.
CrowdGuard~\cite{crowdguard} uses TEEs for integrity-preserving FL. In each round, data providers evaluate the locally trained models of other data providers on their own data inside TEEs and vote on the poisoned model. Unlike \sys, CrowdGuard assumes TEE confidentiality, making it more vulnerable to side-channel attacks.
Laminator~\cite{laminator} and Atlas~\cite{atlas} use TEEs to attest certain properties of centralized ML training and inference. Indeed, the claims produced by \sys{} during FL training could be used by Laminator when performing attested inference. With minor modifications, both Laminator and Atlas could be adapted to use exclaves to remove reliance on TEE confidentiality.

\mypar{Ensuring integrity} \emph{Proof-carrying code}~\cite{necula1997proof} and \emph{model checking}~\cite{clarke1997model} can show that code complies with integrity policies by verifying it against a specification. Since \sys relies on execution integrity, it can be combined with such approaches.
Other techniques enforce \emph{control-flow integrity} during execution: C-FLAT~\cite{abera2016cflat} and LO-FAT~\cite{lo-fat} attest intra-program control flow to ascertain the current execution state of a single node. In contrast, \sys attests task-level data transformations across distributed participants, allowing for verifiable claims about a final trained FL model. While there are similarities (\eg the use of hashes), the attestation objective is different, \eg two FL training runs with identical control-flow attestations could still produce different final models.



\section{Conclusions}
\label{sec:concl}

We described \sys, a new system for executing FL training jobs that enables an auditor to verify claims about the final trained FL model. \sys introduces \emph{exclaves} as a new abstraction for the execution of FL tasks \emph{with integrity only}, which attest all performed data transformations at runtime. Since executed FL tasks are attested through exclave data records~(EDRs), participants cannot deviate from the correct FL training protocol---thus violating model claims---without detection. Exclaves produce EDRs, which can be combined to create an exclave dataflow graph~(EDG) that spans the entire FL job. An auditor uses the EDG to verify end-to-end claims about the FL job. We showed that \sys can be integrated with a state-of-the-art FL framework (NVFlare) to detect a wide range of deviations with low runtime overhead.

\mypar{Acknowledgments} We thank the reviewers and our shepherd for their insightful comments and feedback. Jinnan Guo is supported by the President's PhD Scholarship at Imperial College London.


\bibliographystyle{ACM-Reference-Format}
\bibliography{confidential-fl}

@inproceedings{mo_ppfl_2021,
  title = {{{PPFL}}: Privacy-Preserving Federated Learning with Trusted Execution Environments},
  booktitle = {Proceedings of the 19th {{Annual International Conference}} on {{Mobile Systems}}, {{Applications}}, and {{Services}}},
  author = {Mo, Fan and Haddadi, Hamed and Katevas, Kleomenis and Marin, Eduard and Perino, Diego and Kourtellis, Nicolas},
  year = {2021},
  month = jun,
  pages = {94--108},
  publisher = {{ACM}},
  doi = {10.1145/3458864.3466628},
  url = {https://dl.acm.org/doi/10.1145/3458864.3466628},
  isbn = {978-1-4503-8443-8}
}

@article{crowdguard,
  title={{CrowdGuard}: Federated backdoor detection in federated learning},
  author={Rieger, Phillip and Krau{\ss}, Torsten and Miettinen, Markus and Dmitrienko, Alexandra and Sadeghi, Ahmad-Reza},
  journal={arXiv preprint arXiv:2210.07714},
  year={2022}
}

@inproceedings{mcmahan2017communication,
  title={Communication-efficient learning of deep networks from decentralized data},
  author={McMahan, Brendan and Moore, Eider and Ramage, Daniel and Hampson, Seth and y Arcas, Blaise Aguera},
  booktitle={Artificial intelligence and statistics},
  pages={1273--1282},
  year={2017},
  organization={PMLR}
}

@article{kairouz2021advances,
  title={Advances and open problems in federated learning},
  author={Kairouz, Peter and McMahan, H Brendan and Avent, Brendan and Bellet, Aur{\'e}lien and Bennis, Mehdi and Bhagoji, Arjun Nitin and Bonawitz, Kallista and Charles, Zachary and Cormode, Graham and Cummings, Rachel and others},
  journal={Foundations and Trends{\textregistered} in Machine Learning},
  volume={14},
  number={1--2},
  pages={1--210},
  year={2021},
  publisher={Now Publishers, Inc.}
}

@article{jere2020taxonomy,
  title={A taxonomy of attacks on federated learning},
  author={Jere, Malhar S and Farnan, Tyler and Koushanfar, Farinaz},
  journal={IEEE Security \& Privacy},
  volume={19},
  number={2},
  pages={20--28},
  year={2021},
  publisher={IEEE}
}

@article{dwork2014algorithmic,
  title={The algorithmic foundations of differential privacy},
  author={Dwork, Cynthia and Roth, Aaron and others},
  journal={Foundations and Trends{\textregistered} in Theoretical Computer Science},
  volume={9},
  number={3--4},
  pages={211--407},
  year={2014},
  publisher={Now Publishers, Inc.}
}

@article{hard2018federated,
  title={Federated learning for mobile keyboard prediction},
  author={Hard, Andrew and Rao, Kanishka and Mathews, Rajiv and Ramaswamy, Swaroop and Beaufays, Fran{\c{c}}oise and Augenstein, Sean and Eichner, Hubert and Kiddon, Chlo{\'e} and Ramage, Daniel},
  journal={arXiv preprint arXiv:1811.03604},
  year={2018}
}

@article{ramaswamy2019federated,
  title={Federated learning for emoji prediction in a mobile keyboard},
  author={Ramaswamy, Swaroop and Mathews, Rajiv and Rao, Kanishka and Beaufays, Fran{\c{c}}oise},
  journal={arXiv preprint arXiv:1906.04329},
  year={2019}
}

@inproceedings{yang2019ffd,
  title={{FFD}: A federated learning based method for credit card fraud detection},
  author={Yang, Wensi and Zhang, Yuhang and Ye, Kejiang and Li, Li and Xu, Cheng-Zhong},
  booktitle={Big Data--BigData 2019: 8th International Congress, Held as Part of the Services Conference Federation, SCF 2019, San Diego, CA, USA, June 25--30, 2019, Proceedings 8},
  pages={18--32},
  year={2019},
  organization={Springer}
}

@article{roth2022nvidia,
  title={{NVIDIA FLARE}: Federated learning from simulation to real-world},
  author={Roth, Holger R and Cheng, Yan and Wen, Yuhong and Yang, Isaac and Xu, Ziyue and Hsieh, Yuan-Ting and Kersten, Kristopher and Harouni, Ahmed and Zhao, Can and Lu, Kevin and others},
  journal={arXiv preprint arXiv:2210.13291},
  year={2022}
}

@inproceedings{tolpegin2020data,
  title={Data poisoning attacks against federated learning systems},
  author={Tolpegin, Vale and Truex, Stacey and Gursoy, Mehmet Emre and Liu, Ling},
  booktitle={Computer Security--ESORICS 2020: 25th European Symposium on Research in Computer Security, ESORICS 2020, Guildford, UK, September 14--18, 2020, Proceedings, Part I 25},
  pages={480--501},
  year={2020},
  organization={Springer}
}

@inproceedings{arnautov2016scone,
  title={{SCONE}: Secure {Linux} containers with {Intel} {SGX}},
  author={Arnautov, Sergei and Trach, Bohdan and Gregor, Franz and Knauth, Thomas and Martin, Andre and Priebe, Christian and Lind, Joshua and Muthukumaran, Divya and O'Keeffe, Dan and Stillwell, Mark L and others},
  booktitle={12th USENIX Symposium on Operating Systems Design and Implementation (OSDI 16)},
  pages={689--703},
  year={2016}
}

@article{guo2024trustworthy,
  title={Trustworthy AI Using Confidential Federated Learning},
  author={Guo, Jinnan and Pietzuch, Peter and Paverd, Andrew and Vaswani, Kapil},
  journal={Communications of the ACM},
  volume = {67},
  number = {9},
  pages = {48–53},
  year={2024},
  publisher={ACM New York, NY}
}

@inproceedings{sun2020mobilebert,
  title={MobileBERT: a Compact Task-Agnostic BERT for Resource-Limited Devices},
  author={Sun, Zhiqing and Yu, Hongkun and Song, Xiaodan and Liu, Renjie and Yang, Yiming and Zhou, Denny},
  booktitle={Proceedings of the 58th Annual Meeting of the Association for Computational Linguistics},
  pages={2158--2170},
  year={2020}
}

@inproceedings{devlin2019bert,
  title={{BERT}: Pre-training of deep bidirectional transformers for language understanding},
  author={Devlin, Jacob and Chang, Ming-Wei and Lee, Kenton and Toutanova, Kristina},
  booktitle={Proceedings of the 2019 conference of the North American chapter of the association for computational linguistics: human language technologies, volume 1 (long and short papers)},
  pages={4171--4186},
  year={2019}
}

@article{dougan2014ncbi,
  title={NCBI disease corpus: a resource for disease name recognition and concept normalization},
  author={Do{\u{g}}an, Rezarta Islamaj and Leaman, Robert and Lu, Zhiyong},
  journal={Journal of biomedical informatics},
  volume={47},
  pages={1--10},
  year={2014},
  publisher={Elsevier}
}

@article{krizhevsky2009learning,
  title={Learning multiple layers of features from tiny images},
  author={Krizhevsky, Alex and Hinton, Geoffrey and others},
  year={2009},
  publisher={Toronto, ON, Canada}
}

@article{sanh2019distilbert,
  title={DistilBERT, a distilled version of BERT: Smaller, faster, cheaper and lighter},
  author={Sanh, Victor and Debut, L and Chaumond, J and Wolf, T},
  journal={arXiv preprint arXiv:1910.01108},
  year={2019}
}

@article{radford2019language,
  title={Language models are unsupervised multitask learners},
  author={Radford, Alec and Wu, Jeffrey and Child, Rewon and Luan, David and Amodei, Dario and Sutskever, Ilya and others},
  journal={OpenAI blog},
  year={2019}
}

@article{merity2016pointer,
  title={Pointer sentinel mixture models},
  author={Merity, Stephen and Xiong, Caiming and Bradbury, James and Socher, Richard},
  journal={arXiv preprint arXiv:1609.07843},
  year={2016}
}

@article{lecun1998gradient,
  title={Gradient-based learning applied to document recognition},
  author={LeCun, Yann and Bottou, L{\'e}on and Bengio, Yoshua and Haffner, Patrick},
  journal={Proceedings of the IEEE},
  volume={86},
  number={11},
  pages={2278--2324},
  year={1998},
  publisher={Ieee}
}

@article{simonyan2014very,
  title={Very deep convolutional networks for large-scale image recognition},
  author={Simonyan, Karen and Zisserman, Andrew},
  journal={arXiv preprint arXiv:1409.1556},
  year={2014}
}

@inproceedings{li2021cipherleaks,
  title={{CIPHERLEAKS}: Breaking Constant-time Cryptography on {AMD} {SEV} via the Ciphertext Side Channel},
  author={Li, Mengyuan and Zhang, Yinqian and Wang, Huibo and Li, Kang and Cheng, Yueqiang},
  booktitle={30th USENIX Security Symposium (USENIX Security 21)},
  pages={717--732},
  year={2021}
}

@inproceedings{li2022systematic,
  title={A systematic look at ciphertext side channels on AMD SEV-SNP},
  author={Li, Mengyuan and Wilke, Luca and Wichelmann, Jan and Eisenbarth, Thomas and Teodorescu, Radu and Zhang, Yinqian},
  booktitle={2022 IEEE Symposium on Security and Privacy (SP)},
  pages={337--351},
  year={2022},
  organization={IEEE}
}

@inproceedings{wang2023pwrleak,
  title={PwrLeak: Exploiting Power Reporting Interface for Side-Channel Attacks on AMD SEV},
  author={Wang, Wubing and Li, Mengyuan and Zhang, Yinqian and Lin, Zhiqiang},
  booktitle={International Conference on Detection of Intrusions and Malware, and Vulnerability Assessment},
  pages={46--66},
  year={2023},
  organization={Springer}
}

@article{sev2020strengthening,
  title={Strengthening VM isolation with integrity protection and more},
  author={AMD},
  year={2020}
}

@inproceedings{bonawitz2017secure,
  title={Practical secure aggregation for privacy-preserving machine learning},
  author={Bonawitz, Keith and Ivanov, Vladimir and Kreuter, Ben and Marcedone, Antonio and McMahan, H Brendan and Patel, Sarvar and Ramage, Daniel and Segal, Aaron and Seth, Karn},
  booktitle={Proceedings of the 2017 ACM SIGSAC Conference on Computer and Communications Security},
  pages={1175--1191},
  year={2017}
}

@INPROCEEDINGS{mishra24tdx,
  author={Witharana, Hasini and Chatterjee, Debapriya and Mishra, Prabhat},
  booktitle={2024 IEEE International Symposium on Hardware Oriented Security and Trust (HOST)}, 
  title={Verifying Memory Confidentiality and Integrity of Intel TDX Trusted Execution Environments}, 
  year={2024},
  volume={},
  number={},
  pages={44-54},
  doi={10.1109/HOST55342.2024.10545349}}

@misc{cca24arm,
  author={ARM},
  url = {https://www.arm.com/architecture/security-features/arm-confidential-compute-architecture},
  title = {Arm Confidential Compute Architecture},
  year = {2024}
}

@article{li2023enabling,
  title={Enabling realms with the arm confidential compute architecture},
  author={Li, Xupeng and Li, Xuheng and Dall, Christoffer and Gu, Ronghui and Nieh, Jason and Sait, Yousuf and Stockwell, Gareth},
  journal={USENIX; login},
  year={2023}
}

@article{dhanuskodi2023creating,
  title={Creating the First Confidential GPUs: The team at NVIDIA brings confidentiality and integrity to user code and data for accelerated computing.},
  author={Dhanuskodi, Gobikrishna and Guha, Sudeshna and Krishnan, Vidhya and Manjunatha, Aruna and O'Connor, Michael and Nertney, Rob and Rogers, Phil},
  journal={Queue},
  volume={21},
  number={4},
  pages={68--93},
  year={2023},
  publisher={ACM New York, NY, USA}
}

@inproceedings{vaswani2023confidential,
  title={Confidential Computing within an {AI} Accelerator},
  author={Vaswani, Kapil and Volos, Stavros and Fournet, C{\'e}dric and Diaz, Antonio Nino and Gordon, Ken and Vembu, Balaji and Webster, Sam and Chisnall, David and Kulkarni, Saurabh and Cunningham, Graham and others},
  booktitle={2023 USENIX Annual Technical Conference (USENIX ATC 23)},
  pages={501--518},
  year={2023}
}

@inproceedings{foreshadow,
  title={Foreshadow: Extracting the keys to the {Intel} {SGX} kingdom with transient {Out-of-Order} execution},
  author={Van Bulck, Jo and Minkin, Marina and Weisse, Ofir and Genkin, Daniel and Kasikci, Baris and Piessens, Frank and Silberstein, Mark and Wenisch, Thomas F and Yarom, Yuval and Strackx, Raoul},
  booktitle={27th USENIX Security Symposium (USENIX Security 18)},
  pages={991--1008},
  year={2018}
}

@inproceedings{spectre,
  title={Spectre Attacks: Exploiting Speculative Execution},
  author={Kocher, Paul and Horn, Jann and Fogh, Anders and Genkin, Daniel and Gruss, Daniel and Haas, Werner and Hamburg, Mike and Lipp, Moritz and Mangard, Stefan and Prescher, Thomas and others},
  booktitle={2019 IEEE Symposium on Security and Privacy (SP)},
  pages={1--19},
  year={2019},
  organization={IEEE Computer Society}
}

@article{cheng2023intel,
  title={Intel {TDX} demystified: A top-down approach},
  author={Cheng, Pau-Chen and Ozga, Wojciech and Valdez, Enriquillo and Ahmed, Salman and Gu, Zhongshu and Jamjoom, Hani and Franke, Hubertus and Bottomley, James},
  journal={ACM Computing Surveys},
  volume={56},
  number={9},
  pages={1--33},
  year={2024},
  publisher={ACM New York, NY}
}

@misc{gpublog,
      title={Announcing {Azure Confidential VMs} with {NVIDIA H100} Tensor Core {GPUs} in Preview}, 
      author={Krishnaprasad Hande},
      year={2023},
      url = {https://techcommunity.microsoft.com/t5/azure-confidential-computing/announcing-azure-confidential-vms-with-nvidia-h100-tensor-core/ba-p/3975389}
}

@misc{h100cc,
      title={Confidential Compute on {NVIDIA} Hopper {H100}}, 
      author={Rob Nertney},
      year={2023},
      url = {https://images.nvidia.com/aem-dam/en-zz/Solutions/data-center/HCC-Whitepaper-v1.0.pdf}
}

@inproceedings{Kuznetsov2021SecureFL,
  author    = {Eugene Kuznetsov and Yitao Chen and Ming Zhao},
  title     = {SecureFL: Privacy Preserving Federated Learning with SGX and TrustZone},
  booktitle = {Proceedings of the Sixth ACM/IEEE Symposium on Edge Computing (SEC)},
  year      = {2021},
  pages     = {55--67},
  doi       = {10.1145/3453142.3491287},
  publisher = {ACM},
  address   = {San Jose, CA, USA}
}

@inproceedings{valerio21gradsec,
author = {Messaoud, Aghiles Ait and Mokhtar, Sonia Ben and Nitu, Vlad and Schiavoni, Valerio},
title = {GradSec: a TEE-based Scheme Against Federated Learning Inference Attacks},
year = {2021},
isbn = {9781450387088},
publisher = {Association for Computing Machinery},
address = {New York, NY, USA},
url = {https://doi.org/10.1145/3477114.3488763},
doi = {10.1145/3477114.3488763},
booktitle = {Proceedings of the First Workshop on Systems Challenges in Reliable and Secure Federated Learning},
pages = {10–12},
numpages = {3},
location = {Virtual Event, Germany},
series = {ResilientFL '21}
}

@article{quoc21secfl,
  author       = {Do Le Quoc and
                  Christof Fetzer},
  title        = {SecFL: Confidential Federated Learning using TEEs},
  journal      = {CoRR},
  volume       = {abs/2110.00981},
  year         = {2021},
  url          = {https://arxiv.org/abs/2110.00981},
  eprinttype    = {arXiv},
  eprint       = {2110.00981},
  bibsource    = {dblp computer science bibliography, https://dblp.org}
}

@article{frazelle2019securing,
  title={Securing the Boot Process: The hardware root of trust},
  author={Frazelle, Jessie},
  journal={Queue},
  volume={17},
  number={6},
  pages={5--21},
  year={2019},
  publisher={ACM New York, NY, USA}
}

@techreport{fossati-tls-attestation-08,
    number =    {draft-fossati-tls-attestation-08},
    type =      {Internet-Draft},
    institution =   {Internet Engineering Task Force},
    publisher = {Internet Engineering Task Force},
    note =      {Work in Progress},
    url =       {https://datatracker.ietf.org/doc/draft-fossati-tls-attestation/08/},
    author =    {Hannes Tschofenig and Yaron Sheffer and Paul Howard and Ionuț Mihalcea and Yogesh Deshpande and Arto Niemi and Thomas Fossati},
    title =     {{Using Attestation in Transport Layer Security (TLS) and Datagram Transport Layer Security (DTLS)}},
    pagetotal = 34,
    year =      2024,
    month =     oct,
    day =       21,
}

@misc{dmverity,
  author = {Linux kernel development community},
  howpublished = {\url{https://www.kernel.org/doc/html/latest/admin-guide/device-mapper/verity.html}},
  title = {dm-verity},
  year = {2024}
}

@misc{svsm,
  author = {Coconut SVSM community},
  howpublished = {\url{https://github.com/coconut-svsm/svsm}},
  title = {Coconut Secure VM Service Module},
  year = {2026}
}

@article{lyu2016understanding,
    title={Understanding the sparse vector technique for differential privacy},
    author={Lyu, Min and Su, Dong and Li, Ninghui},
    journal={arXiv preprint arXiv:1603.01699},
    year={2016}
}

@article{blanchard2017machine,
  title={Machine learning with adversaries: Byzantine tolerant gradient descent},
  author={Blanchard, Peva and El Mhamdi, El Mahdi and Guerraoui, Rachid and Stainer, Julien},
  journal={Advances in neural information processing systems},
  volume={30},
  year={2017}
}

@article{pillutla2022robust,
  title={Robust aggregation for federated learning},
  author={Pillutla, Krishna and Kakade, Sham M and Harchaoui, Zaid},
  journal={IEEE Transactions on Signal Processing},
  volume={70},
  pages={1142--1154},
  year={2022},
  publisher={IEEE}
}

@misc{fio,
  author = {Axboe, Jens},
  howpublished = {\url{https://github.com/axboe/fio}},
  title = {{Flexible I/O Tester (fio)}},
  year = {2026},
  note = {Accessed: 2026-07-06}
}

@book{grpc,
  title={gRPC: up and running: building cloud native applications with Go and Java for Docker and Kubernetes},
  author={Indrasiri, Kasun and Kuruppu, Danesh},
  year={2020},
  publisher={O'Reilly Media}
}

@inproceedings{sharma2019secure,
  title={Secure and efficient federated transfer learning},
  author={Sharma, Shreya and Xing, Chaoping and Liu, Yang and Kang, Yan},
  booktitle={2019 IEEE international conference on big data (Big Data)},
  pages={2569--2576},
  year={2019},
  organization={IEEE}
}

@inproceedings{necula1997proof,
  title={Proof-carrying code},
  author={Necula, George C},
  booktitle={Proceedings of the 24th ACM SIGPLAN-SIGACT symposium on Principles of programming languages},
  pages={106--119},
  year={1997}
}

@inproceedings{clarke1997model,
  title={Model checking},
  author={Clarke, Edmund M},
  booktitle={Foundations of Software Technology and Theoretical Computer Science: 17th Conference Kharagpur, India, December 18--20, 1997 Proceedings 17},
  pages={54--56},
  year={1997},
  organization={Springer}
}

@inproceedings{yin2018byzantine,
  title={Byzantine-robust distributed learning: Towards optimal statistical rates},
  author={Yin, Dong and Chen, Yudong and Ramchandran, Kannan and Bartlett, Peter},
  booktitle={International Conference on Machine Learning},
  pages={5650--5659},
  year={2018},
  organization={{PMLR}}
}

@inproceedings{flshield,
  title={{FLShield}: a validation based federated learning framework to defend against poisoning attacks},
  author={Kabir, Ehsanul and Song, Zeyu and Rashid, Md Rafi Ur and Mehnaz, Shagufta},
  booktitle={2024 IEEE Symposium on Security and Privacy (SP)},
  pages={2572--2590},
  year={2024},
  organization={IEEE}
}

@article{gabrielli2023protecting,
  title={Protecting federated learning from extreme model poisoning attacks via multidimensional time series anomaly detection},
  author={Gabrielli, Edoardo and Belli, Dimitri and Matrullo, Zoe and Miori, Vittorio and Tolomei, Gabriele},
  journal={arXiv preprint arXiv:2303.16668},
  year={2023}
}

@inproceedings{merkletree,
  title={A digital signature based on a conventional encryption function},
  author={Merkle, Ralph C},
  booktitle={Conference on the theory and application of cryptographic techniques},
  pages={369--378},
  year={1987},
  organization={Springer}
}

@inproceedings{abera2016cflat,
author = {Abera, Tigist and Asokan, N. and Davi, Lucas and Ekberg, Jan-Erik and Nyman, Thomas and Paverd, Andrew and Sadeghi, Ahmad-Reza and Tsudik, Gene},
title = {{C-FLAT}: Control-Flow Attestation for Embedded Systems Software},
year = {2016},
isbn = {9781450341394},
publisher = {Association for Computing Machinery},
address = {New York, NY, USA},
url = {https://doi.org/10.1145/2976749.2978358},
doi = {10.1145/2976749.2978358},
booktitle = {Proceedings of the 2016 ACM SIGSAC Conference on Computer and Communications Security},
pages = {743–754},
numpages = {12},
location = {Vienna, Austria},
series = {CCS '16}
}

@article{amd_kds,
  title={Hardware VM Isolation in the Cloud},
  author={Kaplan, David},
  journal={Communications of the ACM},
  volume={67},
  number={1},
  pages={54--59},
  year={2024},
  publisher={ACM New York, NY, USA}
}

@article{a3fl,
  title={{A3FL}: Adversarially adaptive backdoor attacks to federated learning},
  author={Zhang, Hangfan and Jia, Jinyuan and Chen, Jinghui and Lin, Lu and Wu, Dinghao},
  journal={Advances in neural information processing systems},
  volume={36},
  pages={61213--61233},
  year={2023}
}

@inproceedings{keystone,
  title={Keystone: An open framework for architecting trusted execution environments},
  author={Lee, Dayeol and Kohlbrenner, David and Shinde, Shweta and Asanovi{\'c}, Krste and Song, Dawn},
  booktitle={Proceedings of the Fifteenth European Conference on Computer Systems},
  pages={1--16},
  year={2020}
}

@article{openfl,
  title={OpenFL: An open-source framework for Federated Learning},
  author={Reina, G Anthony and Gruzdev, Alexey and Foley, Patrick and Perepelkina, Olga and Sharma, Mansi and Davidyuk, Igor and Trushkin, Ilya and Radionov, Maksim and Mokrov, Aleksandr and Agapov, Dmitry and others},
  journal={arXiv preprint arXiv:2105.06413},
  year={2021}
}

@inproceedings{zombieload,
  title={ZombieLoad: Cross-privilege-boundary data sampling},
  author={Schwarz, Michael and Lipp, Moritz and Moghimi, Daniel and Van Bulck, Jo and Stecklina, Julian and Prescher, Thomas and Gruss, Daniel},
  booktitle={Proceedings of the 2019 ACM SIGSAC Conference on Computer and Communications Security},
  pages={753--768},
  year={2019}
}

@inproceedings{trustzone,
  title={{TrustZone Explained}: Architectural features and use cases},
  author={Ngabonziza, Bernard and Martin, Daniel and Bailey, Anna and Cho, Haehyun and Martin, Sarah},
  booktitle={2016 IEEE 2nd International Conference on Collaboration and Internet Computing (CIC)},
  pages={445--451},
  year={2016},
  organization={IEEE}
}

@inproceedings{modelcard,
  title={Model cards for model reporting},
  author={Mitchell, Margaret and Wu, Simone and Zaldivar, Andrew and Barnes, Parker and Vasserman, Lucy and Hutchinson, Ben and Spitzer, Elena and Raji, Inioluwa Deborah and Gebru, Timnit},
  booktitle={Proceedings of the conference on fairness, accountability, and transparency},
  pages={220--229},
  year={2019}
}

@misc{datasheetsdatasets,
    title={Datasheets for Datasets}, 
    author={Timnit Gebru and Jamie Morgenstern and Briana Vecchione and Jennifer Wortman Vaughan and Hanna Wallach and Hal Daumé III and Kate Crawford},
    year={2021},
    eprint={1803.09010},
    archivePrefix={arXiv},
    primaryClass={cs.DB},
    url={https://arxiv.org/abs/1803.09010}
}

@misc{apertus,
      title={Apertus: Democratizing Open and Compliant LLMs for Global Language Environments}, 
      author={Alejandro Hernández-Cano and Alexander Hägele and Allen Hao Huang and Angelika Romanou and Antoni-Joan Solergibert and others},
      year={2025},
      eprint={2509.14233},
      archivePrefix={arXiv},
      primaryClass={cs.CL},
      url={https://arxiv.org/abs/2509.14233}
}

@misc{gpt5,
  title        = {GPT-5 System Card},
  howpublished = {\url{https://cdn.openai.com/gpt-5-system-card.pdf}},
  author       = {OpenAI},
  year         = {2025},
  note         = {Accessed: 2025-09-22}
}

@misc{EUAIAct,
  title        = {Regulation (EU) 2024/1689 of the European Parliament and of the Council of 13 June 2024 laying down harmonised rules on artificial intelligence (Artificial Intelligence Act)},
  author       = {EU},
  howpublished = {\url{https://eur-lex.europa.eu/legal-content/EN/TXT/?uri=CELEX:32024R1689}},
  year         = {2024},
  note         = {Accessed: 2025-09-22}
}

@misc{vtpm,
  title        = {Virtual {TPMs} in {Azure} confidential {VMs}},
  author       = {Microsoft},
  howpublished = {\url{https://learn.microsoft.com/en-us/azure/confidential-computing/virtual-tpms-in-azure-confidential-vm}},
  year         = {2026},
  note         = {Accessed: 2026-07-07}
}

@inproceedings{jia2021pol,
  title={Proof-of-learning: Definitions and practice},
  author={Jia, Hengrui and Yaghini, Mohammad and Choquette-Choo, Christopher A and Dullerud, Natalie and Thudi, Anvith and Chandrasekaran, Varun and Papernot, Nicolas},
  booktitle={2021 IEEE Symposium on Security and Privacy (SP)},
  pages={1039--1056},
  year={2021},
  organization={IEEE}
}

@inproceedings{abbaszadeh2024zero,
  title={Zero-knowledge proofs of training for deep neural networks},
  author={Abbaszadeh, Kasra and Pappas, Christodoulos and Katz, Jonathan and Papadopoulos, Dimitrios},
  booktitle={Proceedings of the 2024 ACM SIGSAC Conference on Computer and Communications Security},
  pages={4316--4330},
  year={2024}
}

@INPROCEEDINGS{zhang2022adversarialPOL,
  author={Zhang, Rui and Liu, Jian and Ding, Yuan and Wang, Zhibo and Wu, Qingbiao and Ren, Kui},
  booktitle={2022 IEEE Symposium on Security and Privacy (SP)}, 
  title={“Adversarial Examples” for Proof-of-Learning}, 
  year={2022},
  volume={},
  number={},
  pages={1408-1422},
  doi={10.1109/SP46214.2022.9833596}
}

@INPROCEEDINGS{fang2023broken,
  author={Fang, Congyu and Jia, Hengrui and Thudi, Anvith and Yaghini, Mohammad and Choquette-Choo, Christopher A. and Dullerud, Natalie and Chandrasekaran, Varun and Papernot, Nicolas},
  booktitle={2023 IEEE 8th European Symposium on Security and Privacy}, 
  title={Proof-of-Learning is Currently More Broken Than You Think}, 
  year={2023},
  volume={},
  number={},
  pages={797-816},
  doi={10.1109/EuroSP57164.2023.00052}
}

@INPROCEEDINGS {sgp,
author = { Tramer, Florian and Zhang, Fan and Lin, Huang and Hubaux, Jean-Pierre and Juels, Ari and Shi, Elaine },
booktitle = { 2017 IEEE European Symposium on Security and Privacy (EuroS\&P) },
title = {{Sealed-Glass Proofs: Using Transparent Enclaves to Prove and Sell Knowledge}},
year = {2017},
volume = {},
ISSN = {},
pages = {19-34},
doi = {10.1109/EuroSP.2017.28},
url = {https://doi.ieeecomputersociety.org/10.1109/EuroSP.2017.28},
publisher = {IEEE Computer Society},
address = {Los Alamitos, CA, USA},
month =apr}

@article{zkp1,
  title={Why and how {zk-SNARK} works},
  author={Petkus, Maksym},
  journal={arXiv preprint arXiv:1906.07221},
  year={2019}
}

@inproceedings{zkp2,
  title={Experimenting with zero-knowledge proofs of training},
  author={Garg, Sanjam and Goel, Aarushi and Jha, Somesh and Mahloujifar, Saeed and Mahmoody, Mohammad and Policharla, Guru-Vamsi and Wang, Mingyuan},
  booktitle={Proceedings of the 2023 ACM SIGSAC Conference on Computer and Communications Security},
  pages={1880--1894},
  year={2023}
}

@article{granite,
  title={{Granite Guardian}},
  author={Padhi, Inkit and Nagireddy, Manish and Cornacchia, Giandomenico and Chaudhury, Subhajit and Pedapati, Tejaswini and Dognin, Pierre and Murugesan, Keerthiram and Miehling, Erik and Cooper, Mart{\'\i}n Santill{\'a}n and Fraser, Kieran and others},
  journal={arXiv preprint arXiv:2412.07724},
  year={2024}
}

@inproceedings{pow,
  title={Proof-of-learning: A blockchain consensus mechanism based on machine learning competitions},
  author={Bravo-Marquez, Felipe and Reeves, Steve and Ugarte, Martin},
  booktitle={2019 IEEE International Conference on Decentralized Applications and Infrastructures (DAPPCON)},
  pages={119--124},
  year={2019},
  organization={IEEE}
}

@inproceedings{buhren2019insecure,
  title={Insecure until proven updated: analyzing AMD SEV's remote attestation},
  author={Buhren, Robert and Werling, Christian and Seifert, Jean-Pierre},
  booktitle={Proceedings of the 2019 ACM SIGSAC Conference on Computer and Communications Security},
  pages={1087--1099},
  year={2019}
}

@inproceedings{lo-fat,
  title={{LO-FAT}: Low-overhead control flow attestation in hardware},
  author={Dessouky, Ghada and Zeitouni, Shaza and Nyman, Thomas and Paverd, Andrew and Davi, Lucas and Koeberl, Patrick and Asokan, N and Sadeghi, Ahmad-Reza},
  booktitle={Proceedings of the 54th Annual Design Automation Conference 2017},
  pages={1--6},
  year={2017}
}

@inproceedings{tpm,
  title={Design and Implementation of the TPM chip J3210},
  author={Zhang, Huanguo and Qin, Zhongping and Yang, Qi},
  booktitle={2008 Third Asia-Pacific Trusted Infrastructure Technologies Conference},
  pages={72--78},
  year={2008},
  organization={IEEE}
}

@misc{tdxquote,
  title        = {Intel® TDX Module Extension for Quoting},
  howpublished = {\url{https://www.intel.com/content/www/us/en/content-details/874303/intel-tdx-module-extension-for-quoting.html}},
  author       = {Intel},
  year         = {2026},
  note         = {Accessed: 2026-06-13}
}

@article{threats_fl,
  title={Threats to training: A survey of poisoning attacks and defenses on machine learning systems},
  author={Wang, Zhibo and Ma, Jingjing and Wang, Xue and Hu, Jiahui and Qin, Zhan and Ren, Kui},
  journal={ACM Computing Surveys},
  volume={55},
  number={7},
  pages={1--36},
  year={2022},
  publisher={ACM New York, NY}
}

@inproceedings{model_inconsistency,
  title={Eluding secure aggregation in federated learning via model inconsistency},
  author={Pasquini, Dario and Francati, Danilo and Ateniese, Giuseppe},
  booktitle={Proceedings of the 2022 ACM SIGSAC Conference on Computer and Communications Security},
  pages={2429--2443},
  year={2022}
}

@inproceedings{bagia2025close,
  title={A Close Look at RMP Entry Caching and Its Security Implications in SEV-SNP},
  author={Bagia, Alexis and Ulitzsch, Vincent Quentin and Trujillo, Dani{\"e}l and Li, Mengyuan and Yan, Mengjia and Seifert, Jean-Pierre},
  booktitle={Proceedings of the 14th International Workshop on Hardware and Architectural Support for Security and Privacy},
  pages={10--18},
  year={2025}
}

@inproceedings{laminator,
author = {Duddu, Vasisht and J\"{a}rvinen, Oskari and Gunn, Lachlan J. and Asokan, N.},
title = {Laminator: Verifiable ML Property Cards using Hardware-assisted Attestations},
year = {2025},
isbn = {9798400714764},
url = {https://doi.org/10.1145/3714393.3726492},
doi = {10.1145/3714393.3726492},
booktitle = {Proceedings of the Fifteenth ACM Conference on Data and Application Security and Privacy},
series = {CODASPY '25}
}

@INPROCEEDINGS{atlas,
  author={Spoczynski, Marcin and Melara, Marcela S. and Szyller, Sebastian},
  booktitle={2025 IEEE European Symposium on Security and Privacy Workshops (EuroS\&PW)}, 
  title={Atlas: A Framework for ML Lifecycle Provenance \& Transparency}, 
  year={2025},
  doi={10.1109/EuroSPW67616.2025.00058}
}

@article{safetynets,
  title={{SafetyNets}: Verifiable execution of deep neural networks on an untrusted cloud},
  author={Ghodsi, Zahra and Gu, Tianyu and Garg, Siddharth},
  journal={Advances in Neural Information Processing Systems},
  volume={30},
  year={2017}
}

@misc{bitcoin,
  title={Bitcoin: A peer-to-peer electronic cash system},
  author={Nakamoto, Satoshi},
  year={2008},
  howpublished = {\url{https://bitcoin.org/bitcoin.pdf}},
  note         = {Accessed: 2026-08-05}
}

\end{document}